\documentclass[acmtog,screen, nonacm]{acmart}
\usepackage{float}
\usepackage{hyperref}
\usepackage{enumitem}
\usepackage{nicefrac}
\usepackage{algorithm}
\usepackage{algorithmic}
\usepackage{amsmath}
\usepackage{acronym}
\usepackage{listings}
\usepackage{xcolor}
\usepackage{multirow}
\usepackage{subcaption}
\usepackage{bm}
\usepackage{stmaryrd}
\usepackage{wrapfig}

\newcommand{\FF}{\bm{F}}

\newcommand{\II}{\bm{I}}

\newcommand{\MM}{\bm{M}}

\newcommand{\CC}{\bm{C}}
\newcommand{\ff}{\bm{f}}
\newcommand{\XX}{\bm{X}}

\newcommand{\LL}{\bm{L}}

\newcommand{\xx}{\bm{x}}

\renewcommand{\AA}{\bm{A}}

\renewcommand{\CC}{\mathbf{C}}

\setcitestyle{square}

\newacro{cv}[CV]{Computer Vision}
\newacro{xpbd}[XPBD]{eXtended Position-based Dynamics}
\newacro{pbd}[PBD]{Position-based Dynamics}
\newacro{mpm}[MPM]{Material Point Method}
\newacro{fem}[FEM]{Finite Element Method}
\newacro{pbf}[PBF]{Position-based Fluids}
\newacro{xpbi}[XPBI]{eXtended Position-based Inelasticity}
\newacro{xsph}[XSPH]{eXtended Smoothed Particle Hydrodynamics}
\newacro{sph}[SPH]{Smoothed Particle Hydrodynamics}
\newacro{mls}[MLS]{Moving Least Squares}
\newacro{cfl}[CFL]{Courant Friedrichs Lewy}
\newacro{admm}[ADMM]{Alternating Direction Methods of Multipliers}
\newacro{ppc}[PPC]{Particle Per Cell}
\newacro{nacc}[NACC]{Non-Associated Cam-Clay}
\newacro{stvk}[StVK]{St. Venant-Kirchhoff}
\newacro{svd}[SVD]{singular value decomposition}
\newacro{dem}[DEM]{Discrete Element Method}

\begin{document}
\title{XPBI: Position-Based Dynamics with Smoothing Kernels Handles Continuum Inelasticity}

% Authors.

\author{Chang Yu}
\orcid{0009-0000-2613-9885}
\authornote{Both authors contributed equally to this research.}
\email{g1n0st@live.com}
\affiliation{
  \institution{UCLA}
  \city{Los Angeles}
  \state{California}
  \country{USA}
}

\author{Xuan Li}
\orcid{0000-0003-0677-8369}
\authornotemark[1]
\email{xuan.shayne.li@gmail.com}
\affiliation{
  \institution{UCLA}
  \city{Los Angeles}
  \state{California}
  \country{USA}
}

\author{Lei Lan}
\orcid{0009-0002-7626-7580}
\email{lanlei.virhum@gmail.com}
\affiliation{
 \institution{University of Utah}
 \city{Salt Lake City}
 \state{Utah}
 \country{USA}
}

\author{Yin Yang}
\orcid{0000-0001-7645-5931}
\email{yangzzzy@gmail.com}
\affiliation{
  \institution{University of Utah}
  \city{Salt Lake City}
  \state{Utah}
  \country{USA}
}

\author{Chenfanfu Jiang}
\orcid{0000-0003-3506-0583}
\email{chenfanfu.jiang@gmail.com}
\affiliation{
 \institution{UCLA}
 \city{Los Angeles}
 \state{California}
 \country{USA}
 }

\renewcommand{\shortauthors}{Chang Yu, et al.}

\begin{abstract}
\acf{pbd} and its extension, \ac{xpbd}, have been predominantly applied to compliant constrained elastodynamics, with their potential in finite strain (visco-) elastoplasticity remaining underexplored. \ac{xpbd} is often perceived to stand in contrast to other meshless methods, such as the \ac{mpm}. \ac{mpm} is based on discretizing the weak form of governing partial differential equations within a continuum domain, coupled with a hybrid Lagrangian-Eulerian method for tracking deformation gradients. In contrast, \ac{xpbd} formulates specific constraints, whether hard or compliant, to positional degrees of freedom. We revisit this perception by investigating the potential of \ac{xpbd} in handling inelastic materials that are described with classical continuum mechanics-based yield surfaces and elastoplastic flow rules. Our inspiration is that a robust estimation of the velocity gradient is a sufficiently useful key to effectively tracking deformation gradients in \ac{xpbd} simulations. By further incorporating implicit inelastic constitutive relationships, we introduce a plasticity in-the-loop updated Lagrangian augmentation to \ac{xpbd}. This enhancement enables the simulation of elastoplastic, viscoplastic, and granular substances following their standard constitutive laws. We demonstrate the effectiveness of our method through high-resolution and real-time simulations of diverse materials such as snow, sand,  and plasticine, and its integration with standard \ac{xpbd} simulations of cloth and water.
\end{abstract}


\begin{CCSXML}
<ccs2012>
   <concept>
       <concept_id>10010147.10010371.10010352.10010379</concept_id>
       <concept_desc>Computing methodologies~Physical simulation</concept_desc>
       <concept_significance>500</concept_significance>
       </concept>
 </ccs2012>
\end{CCSXML}

\ccsdesc[500]{Computing methodologies~Physical simulation}

\keywords{position-based dynamics, material point method, continuum mechanics, elastoplasticity, viscoplasticity}

\begin{teaserfigure}\includegraphics[width=\textwidth]{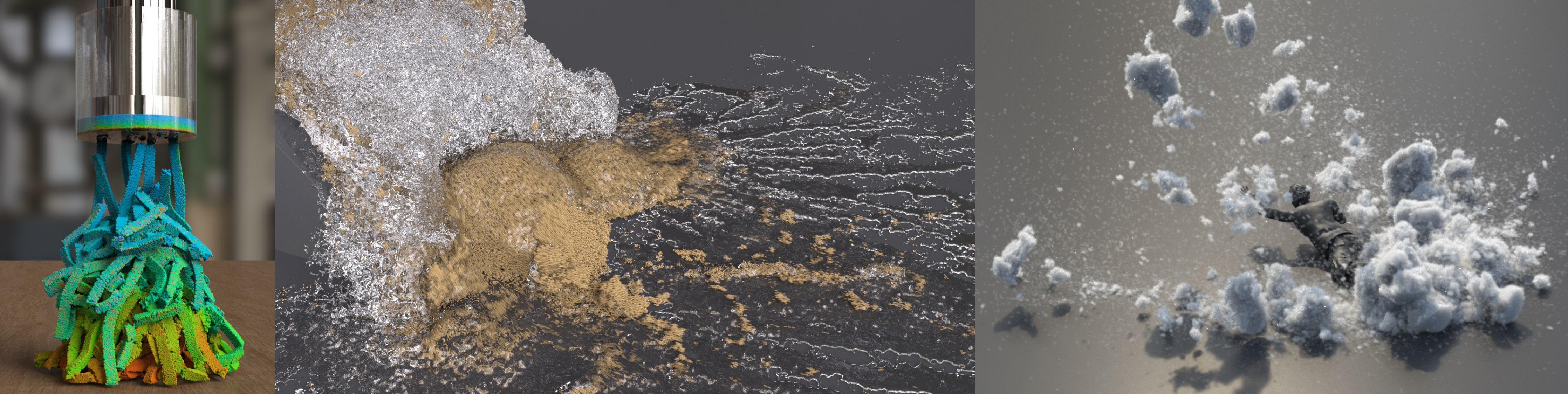}
  \caption{XPBI supports simulating a wide range of classical continuum elastoplastic material models such as Von-Mises plasticine, Drucker-Prager sand, and Cam Clay snow, as well as their interactions with traditional PBD materials such as Position-based Fluid. 
  }
  \label{fig:teaser}
\end{teaserfigure}

\maketitle

\section{Introduction}

\acf{pbd} \cite{muller2007pbd} and its extension \acf{xpbd} \cite{macklin2016xpbd} are widely adopted in compliant constrained dynamics, particularly favored for their performance and simplicity for graphics applications such as rigid bodies \cite{muller2020rigid}, soft bodies \cite{bender2014continuous}, cloth \cite{muller2007pbd}, rods \cite{umetani2015rod} and hair \cite{muller2012hair}. When simulating mesh-based elasticity, it is straightforward to model \ac{xpbd} constraints with FEM hyperelastic energies defined over explicit mesh topology. This allows for effective simulations of elasticity while maintaining the stability and speed of \ac{xpbd}. For example, quadratic energy potentials \cite{chen2023primal} can be formulated using \ac{xpbd}-style constraints.  \citet{macklin2021constraint} and \citet{ton2023parallel} reformulated stable Neo-Hookean \cite{smith2018stable} to demonstrate \ac{xpbd}'s capability in handling nonlinear elasticity.

For inelasticity, on the other hand, two significant challenges emerge. Firstly, topology changes during material splitting and merging introduce great complexity in maintaining a high quality mesh, often necessitating remeshing. Secondly, while there have been explorations that enhances \ac{pbf} \cite{macklin2013pbf} with the conformation tensor \cite{barreiro2017conformation} for viscoelastic fluids, it remains underexplored for \ac{xpbd} to model physically-grounded finite strain (visco-) elastoplastic constitutive laws from classical continuum mechanics, such as von-Mises \cite{Mises1913}, Drucker-Prager  \cite{Drucker1952SoilMA} and Herschel-Bulkley   \cite{Herschel1926KonsistenzmessungenVG} flow rules. Being able to simulate them would greatly improve \ac{xpbd}'s versatility and intuitive controllability of material parameters.

\begin{figure}[t]
\includegraphics[width=1.0\linewidth]{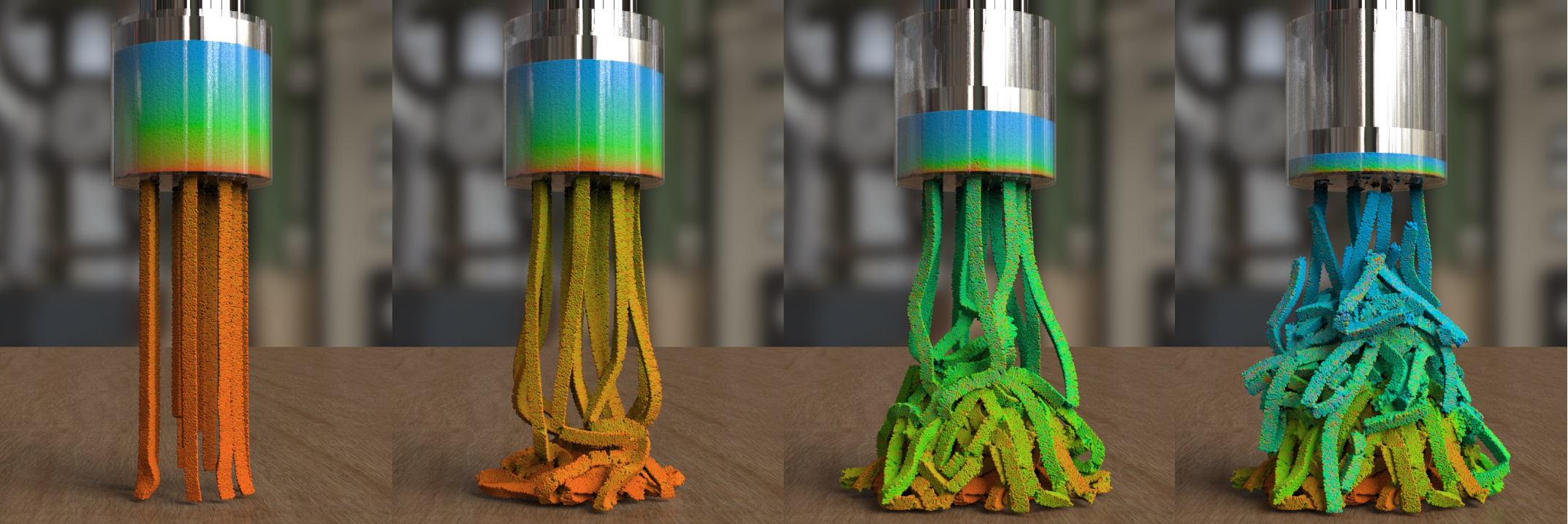}
\caption{\textbf{Noodles.} We simulate noodles modeled using Von Mises plasticity as it is pressed through a cylindrical mold.}
  \label{fig:noodles}
\end{figure}

In comparison with \ac{xpbd}'s success in mesh-based materials, \acf{mpm}'s development in graphics over the past decade has majorly focused on inelastic phenomena with topology change. 
\ac{mpm} is based on the weak form of governing PDEs and employs a hybrid Lagrangian-Eulerian approach for spatial discretization. It handles topology changes, self collision, and finite strain deformation without overhead, and thus has been used for many continuum inelastic phenomena such as snow \cite{stomakhin2013snow}, lava \cite{stomakhin2014augmented}, sand \cite{klar2016sand}, mud \cite{tampubolon2017mixure}, metal \cite{wang2020hot}, foam \cite{ram2015foam}, and fracture \cite{wolper2019fracture}. While substantial progress has been made in these areas, MPM exhibits several notable drawbacks, including excessive numerical dissipation due to particle-grid transfers \cite{jiang2015affine}, artificial stickiness that hampers material separation \cite{fei2021revisiting}, and resolution-dependent gaps between colliding materials \cite{jiang2017anisotropic}. None of these issues are present in \ac{xpbd}. This raises a natural question: \emph{Can we simulate MPM-style phenomena using \ac{xpbd} instead?}

\begin{table}[b]
\caption{XPBI keeps PBD's pure particle nature of the Degrees of Freedom while allowing MPM-style plasticity and granular material modeling through an updated Lagrangian treatment of the deformation and classical continuum mechanics-based elastoplastic flow rules.}
\label{tab:compare-mpm-pbd}
\small{
\begin{tabular}{|c|c|c|c|c|}
\hline
 & \textbf{Lagrangian} & \textbf{DOF} & \textbf{Plasticity} & \textbf{Granular} \\ \hline
\textbf{MPM}    & Updated          & Grid     & Flow Rule & Continuum      \\ \hline
\textbf{PBD} & Total            & Particle & N/A       & Sphere Approx. \\ \hline
\textbf{Ours}   & Total \& Updated & Particle & Flow Rule & Continuum      \\ \hline
\end{tabular}
}
\end{table}

Towards addressing this question, we make an important observation: the primary factor that facilitates the modeling of inelasticity lies not in the hybrid Lagrangian-Eulerian nature of MPM, but rather in its use of an updated Lagrangian formulation for the deformation gradient tensor. 
In particular, one considers the time $n+1$ velocity $\bm{v}^{n+1}$ to be defined over the previous time $n$ domain $\Omega^n$ through the Lagrangian velocity $\bm{V}$ of a particle traced back using the inverse deformation map $\bm{\phi}^{-1}(\xx,t)$:
$\bm{v}^{n+1}(\xx) = \bm{V}(\bm{\phi}^{-1}(\xx,t^n),t^{n+1})$ \cite{jiang2016material}, where $t$ represents the continuous time variable. 
This enables the derivation of the rate form of the deformation gradient $\FF$ given by $\dot{\FF}=(\nabla \bm{v})\FF$, which can be further discretized into $\FF^{n+1} =(\II+\Delta t \nabla \bm{v}^{n+1})\FF^n$, where $\Delta t$ is the time step size, allowing one to track the deformation gradient \emph{without referring to a material space configuration}.

Inspired by this observation, if we can track the deformation gradient tensor $\FF$ using an updated Lagrangian view in \ac{xpbd}, then by treating $\FF$ as a function of \ac{xpbd} degrees of freedom, we can modify \ac{xpbd} to resemble a ``material point'' approach. As detailed in later sections, this task reduces to robustly computing and differentiating the velocity gradient tensor.  We present developments surrounding these ideas by introducing \acf{xpbi}, where the \emph{X} represents not only the incorporation of XPBD augmentation but also the use of velocity as the primal variable. This approach computes the updated Lagrangian deformation gradient using a velocity-based formulation, allowing us to handle various inelastic effects. Using velocities as primary variables allows for direct evaluation of the velocity gradient and particle-wise constraints using interpolation kernels defined at $t^n$, aligning with standard MPM practices, while using positions could introduce uncertainties in updating kernels during implicit iterations. By further incorporating an implicit plasticity treatment and additional stability-enhancing components, our method leverages the efficiency and simplicity of \ac{pbd} while capturing the complex inelastic material responses typically associated with \ac{mpm}; see Table.~\ref{tab:compare-mpm-pbd}. In summary, our contributions include:
\begin{itemize}
    \item An updated Lagrangian augmentation for \ac{xpbd} that tracks meshless deformation gradients and per-particle constraints;
    \item \ac{xpbi}, a fully implicit plasticity-aware algorithm capable of handling continuum mechanics-based elastoplastic/viscoplastic laws;
    \item An investigation for practical stability enhancements, such as XSPH and position correction, and validations of our method with various practical examples.
\end{itemize}

\section{Related Work}

\paragraph{Inelasticity with \ac{pbd}}

\citet{muller2007pbd} introduced \ac{pbd}, which replaces internal forces with positional constraints and produces appealing, stable and real-time simulations. Its first-order convergence was studied by \citet{plunder2023convergence}. \ac{xpbd} \cite{macklin2016xpbd}, an extension of \ac{pbd}, utilizes the compliant-constraint framework \cite{tournier2015constrained} to uniformly handle soft and hard constraints to simulate elasticity.  Our work follows the latest XPBD paradigm \cite{macklin2019small} with substeps.
Other \ac{pbd} materials include rigid body \cite{muller2020rigid}, soft body \cite{bender2014continuous}, cloth \cite{muller2007pbd}, hair \cite{muller2012hair}, elastic rod \cite{umetani2015rod}, sand \cite{macklin2014unified}, fluid \cite{macklin2013pbf} with surface tension \cite{xing2022tension} and their unified couplings \cite{macklin2014unified, francu2017unified, abu2020isph}. We refer to \citet{bender2017survey} for a comprehensive survey.

\begin{figure} [t]
\includegraphics[width=1.0\linewidth]{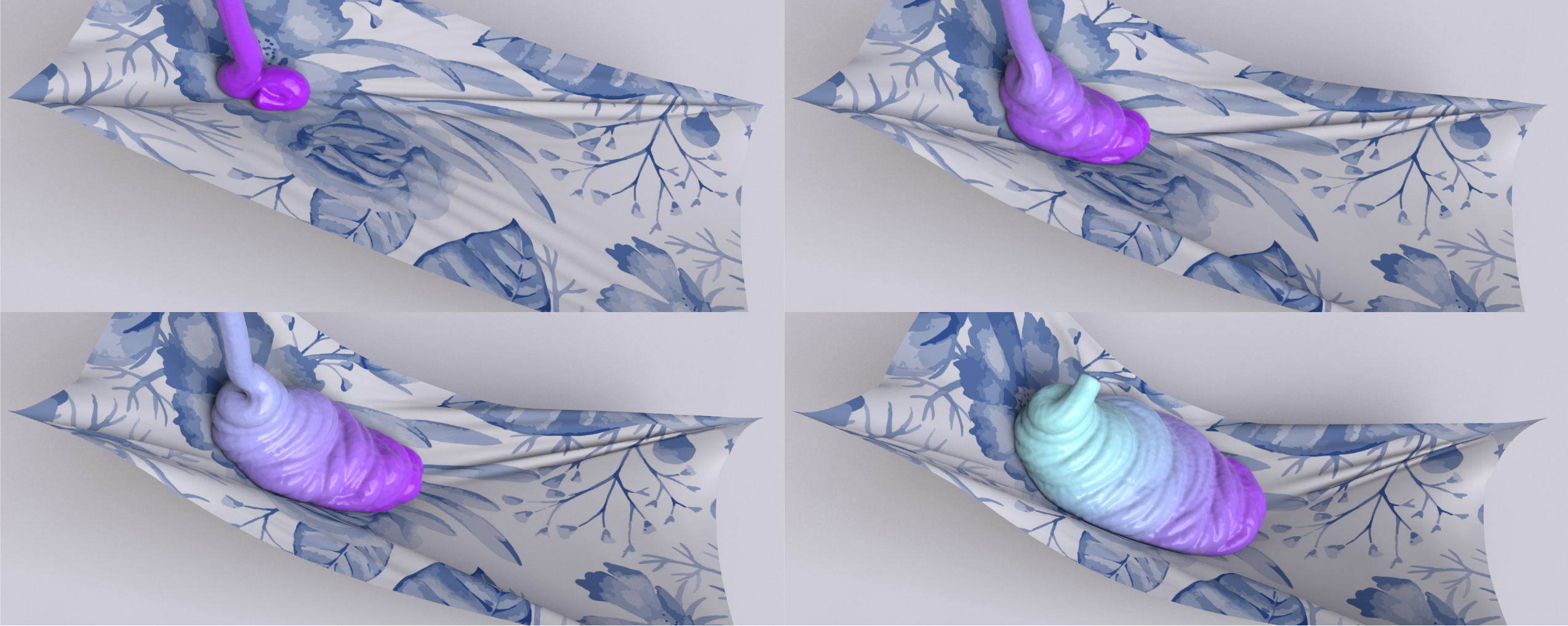}
  \caption{\textbf{Cloth.} XPBI fits into traditional XPBD pipeline and naturally couples updated Lagrangian materials (viscoplastic paint) and mesh-based geometry (cloth).}
  \label{fig:cloth_sequence}
\end{figure}

For \emph{continuum} materials, \citet{bender2014continuous} defined a constraint for the elastic strain energy. \citet{muller2015strain} constrained the strain tensor directly instead. \citet{macklin2021constraint} reformulated stable Neo-Hookean using \ac{xpbd}. Plastic deformation and fracture can be modeled by shape matching \cite{chentanez2016real, jones2016ductile, falkenstein2017reclustering},  prioritizing efficiency over accuracy. \citet{macklin2014unified} simulated sand as colliding spheres with friction. SideFX Houdini's Vellum \ac{pbd} solver further added spring-like cohesion for snow.  Without further utilizing continuum mechanics-based inelastic models, these approaches have limited mechanical intuition and physical parameter controllability. A step forward was proposed by \citet{barreiro2017conformation}, which enhances \ac{pbf} with the conformation tensor for viscoelastic fluids. Nevertheless, it does not incorporate finite strain continuum mechanics, limiting its suitability in modeling general elastoplastic and viscoplastic laws.

\paragraph{Inelasticity with \ac{mpm}} \ac{mpm} was introduced by \citet{sulsky1994particle} as a hybrid Lagrangian/Eulerian approach for solids. Since its adoption in graphics \cite{hegemann2013level, stomakhin2013snow}, \ac{mpm} has gained significant attentions for its automatic topology change and material versatility. Snow plasticity was first done by projecting principal stretches \cite{stomakhin2013snow}. Phase change was modeled by \shortcite{stomakhin2014augmented} through a dilational/deviatoric splitting of the constitutive model. \citet{yue2015continuum} adopted Herschel-Bulkley viscoplasticity for foam.  \citet{ram2015foam} used the Oldroyd-B model for viscoelastic fluids. \citet{fei2019multi} developed an analytical plastic flow approach for shear-dependent liquids. Following Drucker-Prager yield criterion, \citet{klar2016sand, daviet2016semi} modeled sand as continuum granular materials and  \citet{tampubolon2017mixure} further added wetting.  \citet{wolper2019fracture,wolper2020anisompm} captured dynamic fracture using \ac{nacc} plasticity and damage mechanics. Advocating implicit integrators, stiff plastic materials like metal was simulated with Newton-Krylov MPM \cite{wang2020hot}, while \citet{fang2019silly} used \ac{admm} for viscoelasticity and elastoplasticity and \citet{li2022energetically} proposed a variational implicit inelasticity formulation.

\paragraph{Inelasticity with Other Discretizations}

\ac{sph} was originally developed for simulating incompressible flow. \citet{clavet2005particle} added dynamic-length springs for viscoelasticity. \citet{jones2014deformation} and \citet{muller2004point} solved \ac{mls} for elastoplasticity. \citet{gerszewski2009point} first used deformation gradient tensor with multiplicative elastoplastic decomposition in SPH. \citet{alduan2011sph} and \citet{yang2017unified} modeled granular materials based on Drucker Prager yielding.  \citet{takahashi2015implicit} used an implicit SPH formulation to simulate viscous fluids. \citet{gissler2020implicit} developed an implicit \ac{sph} snow solver similarly to \citet{stomakhin2013snow}'s MPM treatment. Using power diagram-based particle-in-cell  \cite{qu2022power} and \ac{mls}-\ac{mpm} \cite{hu2018moving}, power plastics \cite{qu2023power} simulated inelastic flow with an \ac{xpbd}-style Gauss-Seidel solver. Peridynamics \cite{silling2000reformulation} defines pairwise forces and integrates  particle interactions. \citet{he2017projective} combined peridynamics with projective dynamics \cite{bouaziz2014projective} and modeled Drucker-Prager plasticity. \citet{chen2018peridynamics} used isotropic linear elasticity with plasticity and simulated fracture.

\begin{figure}[b]
\includegraphics[width=1.0\linewidth]{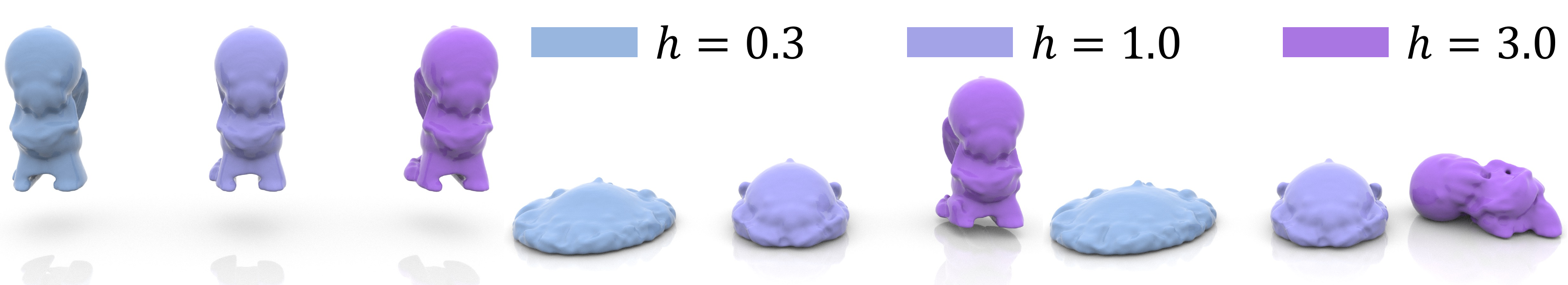}
  \caption{XPBI simulates Hershel-Bulkley shear thinning ($h=0.3$), viscoplastic ($h=1.0$), and shear thickening materials ($h=3.0$), where $h$ controls a power law flow rate detailed in \cite{yue2015continuum}.}
  \label{fig:shear_thickening}
\end{figure}

\section{Method}

We start with briefly reviewing \ac{xpbd} \cite{macklin2016xpbd}, which lets hyperelasticity be governed by Newton's equations of motion through a potential $U(\xx)$: \(\MM \ddot{\xx} = - \nabla U^T (\xx) \), where $\MM$ is the mass matrix and \(\xx=[x_1, x_2, ..., x_p]^T\) is the unknown position states. XPBD assumes $U(\xx)$ can be further expressed as $U(\xx)=\frac{1}{2} \CC(\xx)^T \bm{a}^{-1} \CC(\xx)$, where $\bm{C} = [C_1(\xx), C_2(\xx), ..., C_m(\xx)]^T$ contains $m$ constraints, and $\bm{a}$ is a diagonal compliance matrix. The elastic internal force $\bm{f}$ and Lagrange multiplier $\bm{\lambda}$ are then shown to be 
\begin{align}
\bm{f} &= -\nabla \CC(\xx)^T \bm{a}^{-1} \CC(\xx), \\
\bm{\lambda} &= - \tilde{\bm{a}}^{-1} \CC(\xx).
\end{align}
where $\tilde{\bm{a}} = \frac{\bm{a}}{\Delta t^2}$ and $\Delta t$ is the time step size. 

\subsection{Rewriting StVK Elasticity as Constraints}
\label{sec:stvk}
For modeling elasticity, we adopt the \ac{stvk} model with Hencky strains. As in \citet{klar2016sand}, \citet{gao2017adaptive}, the advantage of this choice is for math/code simplicity and runtime efficiency – it allows return mapping to have analytical solutions for certain plastic flows, eliminating the need for numerical solutions. The elastoplastic behavior of isotropic materials is characterized in the principal stretch space $\bm{\Sigma}$ via \ac{svd} \cite{stomakhin2012energetically} of the deformation gradient $\FF$. The element's total potential energy $\Phi$ can be expressed as $\Phi=V^0\Psi$, where $V^0$ is an element's rest volume and $\Psi$ is the energy density, assuming piecewise constant element deformations, i.e., one particle has one deformation gradient. For \ac{stvk} we have energy density $\Psi$
\begin{align}
\Psi = \mu \text{tr}\left(\log\left(\bm{\Sigma}\right)^2\right) + \frac{\lambda}{2} \left(\text{tr}\left(\log\left(\bm{\Sigma}\right)\right)\right)^2,
\end{align}
where $\mu$ and $\lambda$ are the Lam\'{e} parameters.

To convert $\Phi$ into constraints, our first option is to separately handle each term. As done by \citet{macklin2021constraint} for Neo-Hookean, by utilizing $\Phi = \frac{1}{2} \frac{1}{\alpha} C^2$, we could define two constraints for the $\mu$ and $\lambda$ term respectively as
\begin{align}
\alpha_{\mu} = 1/ (2 \mu V^0), &\text{ and} \ C_{\mu} =  \sqrt{\text{tr} \left(\log\left(\bm{\Sigma}\right)^2 \right)}; \\
\alpha_{\lambda} = 1/(\lambda V^0), &\text{ and} \ C_{\lambda} = \text{tr}\left(\log\left(\bm{\Sigma}\right)\right).
\end{align}
Alternatively, as done by \citet{qu2023power} on power diagrams, we can absorb Lam\'{e} parameters from $\alpha$ to $C$ and define a single $\FF$-dependent constraint through $\Phi = V^0 \Psi(\FF) = \frac{1}{2 \alpha} C(\FF)^2$ to achieve 
\begin{align}
\alpha = {1}/{V^0}, \text{ and} \ C(\FF) = \sqrt{2\Psi(\FF)}, 
\label{eqn:ourCofF}
\end{align}
which clearly halves the number of constraints by allowing more nonlinearity in each $C(\FF(\xx))$. In practice we find both options effective, and adopt the single-constraint version for efficiency.

\begin{figure} [t]
\includegraphics[width=1.0\linewidth]{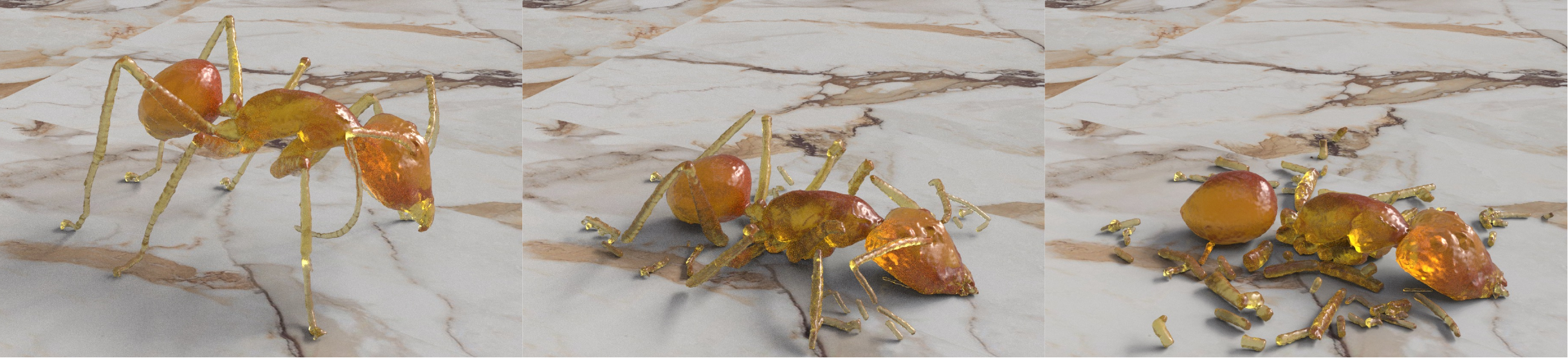}
\caption{\textbf{Candy Camponotus.} We simulate the brittle fracture of a candy shaped like a camponotus falling onto the ground.}
  \label{fig:camponotus}
\end{figure}

\subsection{Gradient is All You Need}
\label{sec:gradient}

\begin{figure}[t]
\centering
\includegraphics[width=1.0\linewidth]{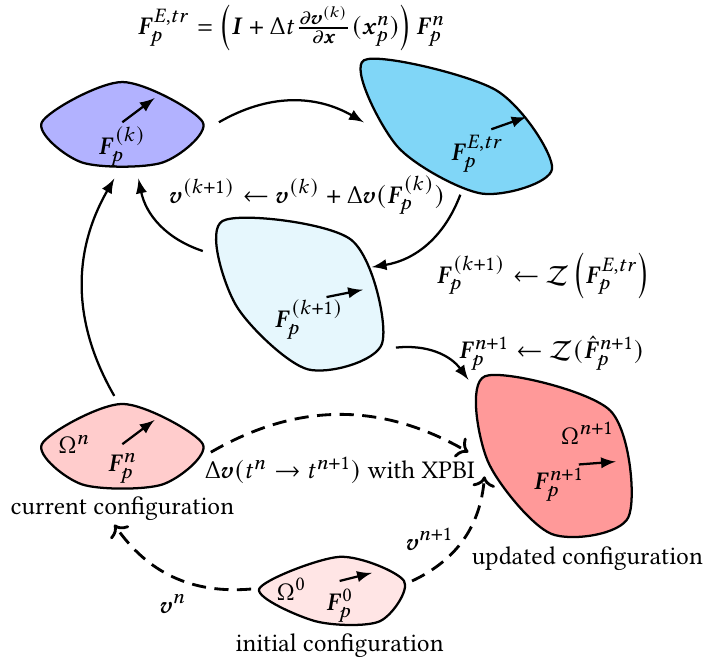}
\caption{\textbf{Deformation Gradient Evolution.} The dotted line (\textbf{bottom}) illustrates the evolution of the deformation gradient in the updated Lagrangian view, transitioning from $\FF_p^0$ (initial configuration) to $\FF_p^{n+1}$ (updated configuration), with $\FF_p^{n}$ (current configuration) serving as the reference state. This facilitates tracking large deformations. The solid line loop (\textbf{top}) depicts iterations of our XPBI algorithm to simulate $t^n \rightarrow t^{n+1}$, alternating between an \ac{xpbd} iteration and a fixed point iteration. During iteration $k$, $\FF_p^{E, tr}$ is first estimated based on the current gradient of $\bm{v}^{(k)}$ (\textsection~\ref{sec:gradient}). Then, plasticity is applied through projection to obtain $\FF_p^{(k+1)}$ (\textsection~\ref{sec:implicit_plasticity}), and finally, $\bm{v}^{(k+1)}$ is updated by solving constraints (\textsection~\ref{sec:algorithm_overview}).}
\label{fig:potato}
\end{figure}

Unlike mesh-based representations \cite{macklin2021constraint} which use linear FEM to model the deformation map $\xx=\bm{\phi}(\XX, t)$ and compute deformation gradient $\FF=\partial \bm{\phi}(\XX, t)/\partial{\XX}$ using an undeformed reference state $\XX \in \Omega^0$, meshless inelastic materials cannot utilize simplex elements due to extreme deformation. We follow \citet{jiang2016material} and \citet{gissler2020implicit} to derive the time rate of the deformation gradient:
\begin{align}
\frac{\partial}{\partial t} \FF(\XX, t) &= \frac{\partial}{\partial t} \frac{\partial \bm{\phi}}{\partial \XX} (\XX, t)  =  \frac{\partial \bm{v}}{\partial \xx}(\bm{\phi}(\XX, t), t) {\frac{\partial \bm{\phi}}{\partial \XX}(\XX, t)},
\end{align}
with $\bm{V}(\XX, t) = \partial \bm{\phi}(\XX, t) / \partial t $ being the Lagrangian velocity whose Eulerian counterpart is $\bm{v}(\xx, t)=\bm{V}(\bm{\phi}^{-1}(\xx, t), t)$. With time discretization from $t^n$ to $t^{n+1}$ and the assumption that the velocity $\bm{v}$ at time $t^{n+1}$ being $\bm{v}^{n+1}(\xx)$ for $\xx \in \Omega^n$, we have
\begin{align}
\frac{\partial}{\partial t} \FF(\XX, t^{n+1}) =  \frac{\partial \bm{v}^{n+1}}{\partial \xx}(\bm{\phi}(\XX, t^n)) \FF(\XX, t^n).
\label{eq:updated_lagrangian0}
\end{align}
Taking 
$\frac{\partial}{\partial t} \FF_p(\XX_p, t^{n+1}) \approx ({\FF^{n+1}_p - \FF^{n}_p})/{\Delta t}$ for a particle $\XX_p$ we get
\begin{align}
\FF_p^{n+1} = \FF_p^{n} + \Delta t \frac{\partial \bm{v}^{n+1}}{\partial \xx}(\xx_p^n) \FF_p^n = \left(\II + \Delta t \frac{\partial \bm{v}^{n+1}}{\partial \xx}(\xx_p^n) \right) \FF_p^n
\label{eq:updated_lagrangian}
\end{align}
as the evolution of $\FF_p^{n+1}$ given $\bm{v}^{n+1}$ and $\FF_p^n$. With updated Lagrangian \shortcite{de2020material}, the reference space is thus always $\Omega^n$ and there is no need to store $\Omega^0$; see Fig.~\ref{fig:potato}. Therefore, to express  constraints as functions of positions $C(\FF(\bm{v}^{n+1}(\xx^{n+1})))$, \emph{robustly estimating $\partial \bm{v} / \partial \xx$ ($\forall x\in \Omega^n$) and its derivative is all one needs.}

\begin{figure*}[t]
\includegraphics[width=1.0\linewidth]{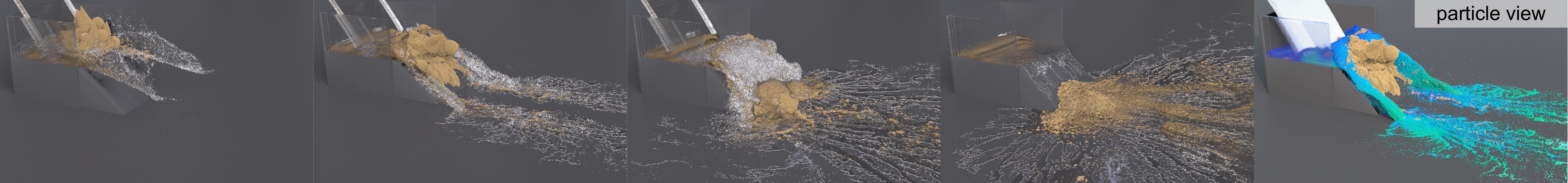}
  \caption{\textbf{Dam Breach.} Our method can be seamlessly coupled with PBF \cite{macklin2013pbf} to simulate sand and water mixture.}
  \label{fig:castle}
\end{figure*}

\begin{figure}[b]
\includegraphics[width=1.0\linewidth]{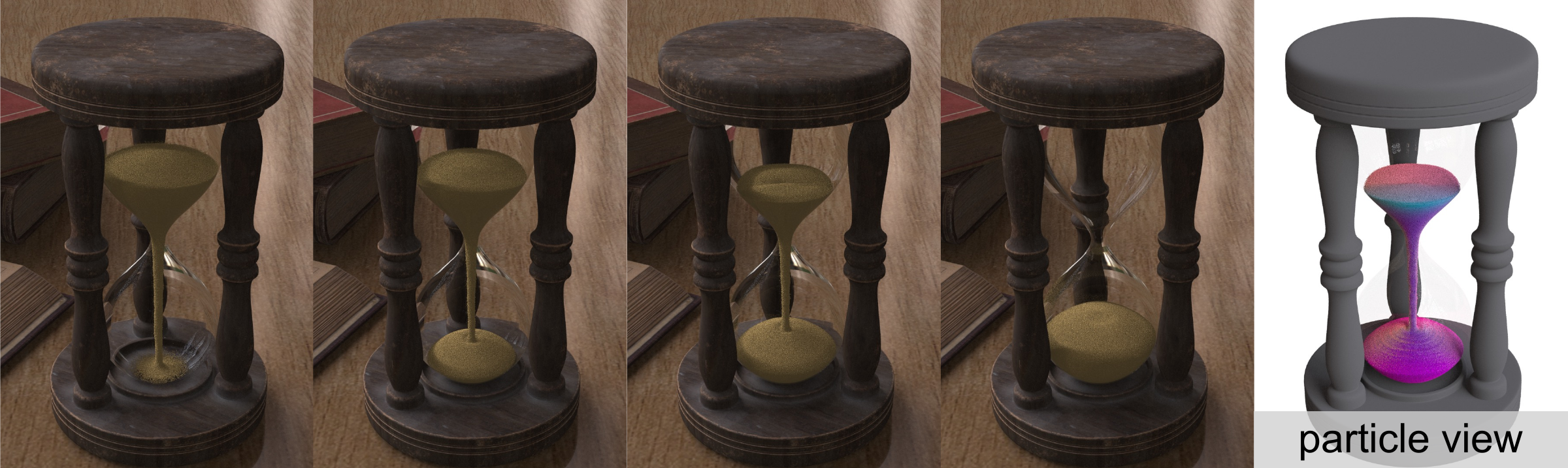}
  \caption{\textbf{Hourglass.} Sand in  hourglass accumulates at the bottom. The material is modeled with Drucker-Prager plasticity \cite{klar2016sand}.}
  \label{fig:hourglass}
\end{figure}

Accurately estimating meshless velocity gradients is generally challenging. Most meshless shape functions require a dense neighborhood to fulfill the kernel's normalization condition. Known as neighborhood deficiency, significant accuracy degradation would occur especially for first-order derivatives (such as velocity gradient) in sparse regions. Kernel gradient correction \cite{bonet1999variational} and the reproducing kernel particle method \cite{liu1995reproducing} are examples of strategies for mitigating this problem. Here, we adopt Wendland kernels \cite{wendland1995piecewise} for the standard SPH kernel $W$ and $\nabla W$ and the reweighting-based kernel gradient correction \cite{bonet1999variational}. The correction matrix $\LL_p$ is defined as
\begin{align}
\LL_p = \left(\sum_b V_b^n \nabla W_b(\xx_p) \otimes (\xx_b-\xx_p)\right)^{-1},
\end{align}
with \ac{svd} based pseudo inverse $\AA^{-1}=\bm{V}\bm{\Sigma}^{-1}\bm{U}^{T}$ to avoid singularities when calculating ill-conditioned matrix inverses for numerical stability, where $V_b^n=V_b^0 \text{det}(\FF_b^n)$ is the time $n$ volume of the neighborhood particle $b$. As in many SPH methods, the correction is indispensable for maintaining simulation stability. See  \citet{westhofen2023comparison} for discussions of similarly viable gradient estimation choices. Our discrete velocity gradient at particle $p$ in $\Omega^n$ is then
\begin{align}
 \frac{\partial \bm{v}}{\partial \xx}(\xx_p^n) &= \sum_{b\neq p} V_b^n (\bm{v}_b-\bm{v}_p) \left(\LL_p \nabla W_b(\xx_p^n)\right)^T,
\end{align}
which is also adopted in \citet{gissler2020implicit}. Combining the gradient estimation with Eq.~\ref{eq:updated_lagrangian} and differentiating per-particle constraint $C_p(\FF_p)$ (Eq.~\ref{eqn:ourCofF}) reveals

\begin{align}
\nabla_{\xx_b}{C_p}|_{b\neq p} &= V_b^n \frac{\partial C_p}{\partial \FF_p} {\FF_p^n}^T  (\LL_p \nabla W_b(\xx_p^n)), \\ \nabla_{\xx_p}C_p &= -\sum_{b\neq p} \nabla_{\xx_b}C_p,
\end{align}
which provides us all necessary constraint derivatives in XPBD.

\subsection{Implicit Plasticity}
\label{sec:implicit_plasticity}

Plasticity in continuum mechanics is typically solved with return mapping, denoted as $\mathcal{Z}(\cdot)$, which adjusts strains according to a plastic flow rule. It projects an elastic predictor $\FF^{E,tr}$ onto the yield surface to ensure an inequality constraint on the stress. 

To make plasticity implicit, we propose to alternate between (1) an \ac{xpbd} iteration with a projected stress and (2) a stress projection. This is essentially a fixed point iteration similarly to  \citet{li2022energetically}:
\begin{align}
\FF_p^{(k+1)} \leftarrow \mathcal{Z}\left(\FF_p^{E, tr}\left(\bm{v}^{(k)}\left(\FF_p^{(k)}\right)\right)\right),
\end{align}
where $\FF_p^{E, tr}$ is the trial elastic deformation gradient and \(\bm{v}^{(k)}(\FF_p^{(k)})\) is the updated velocity by previous $k$ \ac{xpbd} iterations based on \(\FF_p^{(k)}\) in the previous iteration. In contrast to \citet{li2022energetically}'s fixed point iteration on $\mathcal{Z}$ which functions as an independent outer loop of a full Newton optimization, our design establishes a fixed point on 
$\FF$, updating variables directly impacted by the fixed point iteration \emph{within} \ac{xpbd} iterations; see Fig.~\ref{fig:potato} top and Alg.~\ref{alg:cap}. Resultingly, our implicit plasticity treatment introduces negligible extra cost on top of what implicit elasticity already necessitated.

Nonetheless, convergence in fixed-point iterations depends on an initial guess sufficiently close to the solution, among other conditions of the implicit function. In this paper, we do not monitor quantitative plasticity convergence since few \ac{xpbd} iterations are needed for visually plausible results. We do emphasize the importance of implicit plasticity and compare it with a semi-implicit treatment which only applies plasticity at the end of a time step; see \textsection~\ref{sec:evaluation}.

\begin{figure}[b]
\includegraphics[width=1.0\linewidth]{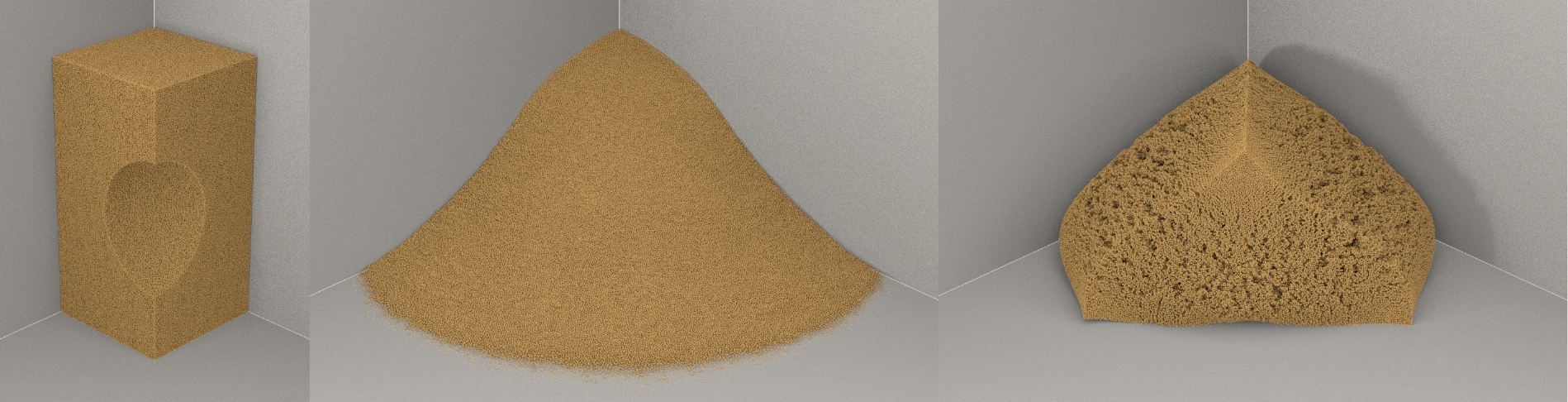}
  \caption{Comparison on notched sand block fall. Inital (\textbf{left}), our fully implicit treatment (\textbf{middle}), and semi-implicit plasticity (\textbf{right}).}
  \label{fig:semi_implicit}
\end{figure}

\section{Algorithm}

Here we detail the XPBI pipeline and its seamlessly integration into existing \ac{xpbd}. Our pseudocode for advancing a time step using velocity-based XPBD is summarized in Alg.~\ref{alg:cap}.

\subsection{Algorithm Overview}
\label{sec:algorithm_overview}

Similarly to MPM, we use material particles to discretize the continuum. Each particle $p$ is governed by a constitutive model-induced constraint $C_p$ (Eq.~\ref{eqn:ourCofF}) and a plastic return mapping operator $\mathcal{Z}$. We use $C_p$ to denote particle-wise inelasticity constraints ($|\{p\}|=N=\# \text{ particles}$) and $C_i$ to denote traditional PBD constraints ($|\{i\}|=M=\# \text{ number of all other constraints}$).

Due to our dependency on velocity gradients, it is more natural to reparametrize XPBD with velocities rather than positions as primary unknown variables. Closely resembling position-based XPBD, we solve for velocities $\bm{v}^{n+1}$ and Lagrange multipliers $\bm{\lambda}^{n+1}$ that satisfies
\begin{align}
\MM(\bm{v}^{n+1}-\tilde{\bm{v}}^{n})-\nabla \bm{C} (\bm{v}^{n+1})^T \bm{\lambda}^{n+1}&=\bm{0}, \\
\bm{C}(\bm{v}^{n+1})+\tilde{\bm{a}}\bm{\lambda}^{n+1} &= \bm{0},
\end{align}
by updating per-particle inelastic constraint $C_p$'s corresponding Lagrangian multiplier $\lambda_p$ with 
\begin{align}
\Delta\lambda_p = \frac{-C_p-\tilde{\alpha}_p\lambda_p}{\sum_{b=1}^N \frac{1}{m_b} |\nabla_{\xx_b}C_p(\xx)|^2 + \tilde{\alpha_p}},
\end{align}
where $\alpha_p=1/V_p^0$ and  $\tilde{\alpha}_p=\alpha_p/ \Delta t^2$. 
The velocity update is given by:
\begin{align}
\Delta\bm{v} = \left(\MM^{-1}\nabla \CC(\xx)^T \Delta \bm{\lambda}\right)/ \Delta t.
\end{align}
The velocities and multipliers are jointly updated by a colored Gauss-Seidel iteration (see \textsection~\ref{sec:colored_gs} and Alg.~\ref{alg:cap}):
\begin{align}
\lambda_p^{(k+1)} \leftarrow \lambda_p^{(k)}+ \Delta \lambda_p,  \quad \bm{v}^{(k+1)} \leftarrow \bm{v}^{(k)}+\Delta \bm{v}.
\end{align}

Note that by selecting velocities as our primary unknown variables, our deformation gradient update (as derived in Eq.~\ref{eq:updated_lagrangian0} and Eq.~\ref{eq:updated_lagrangian}) within the Gauss-Seidel iteration allows us to directly evaluate the velocity gradient and particle-wise constraints in $\Omega^n$ using interpolation kernels defined at $t^n$, aligning with a typical MPM approach \citep{jiang2016material}. Conversely, using positions as primary variables could introduce uncertainties regarding whether to update the kernels during implicit iterations, which is an intriguing area for future exploration. For collisions with standard PBD materials, as illustrated in the loop over $C_i$  in Alg.~\ref{alg:cap}, we follow the standard PBD approach, where constraints are directly evaluated using collision kernels defined by the latest candidate positions during the iterations, ensuring accurate prediction of potential collisions.

\begin{figure} [t]
\includegraphics[width=0.8\linewidth]{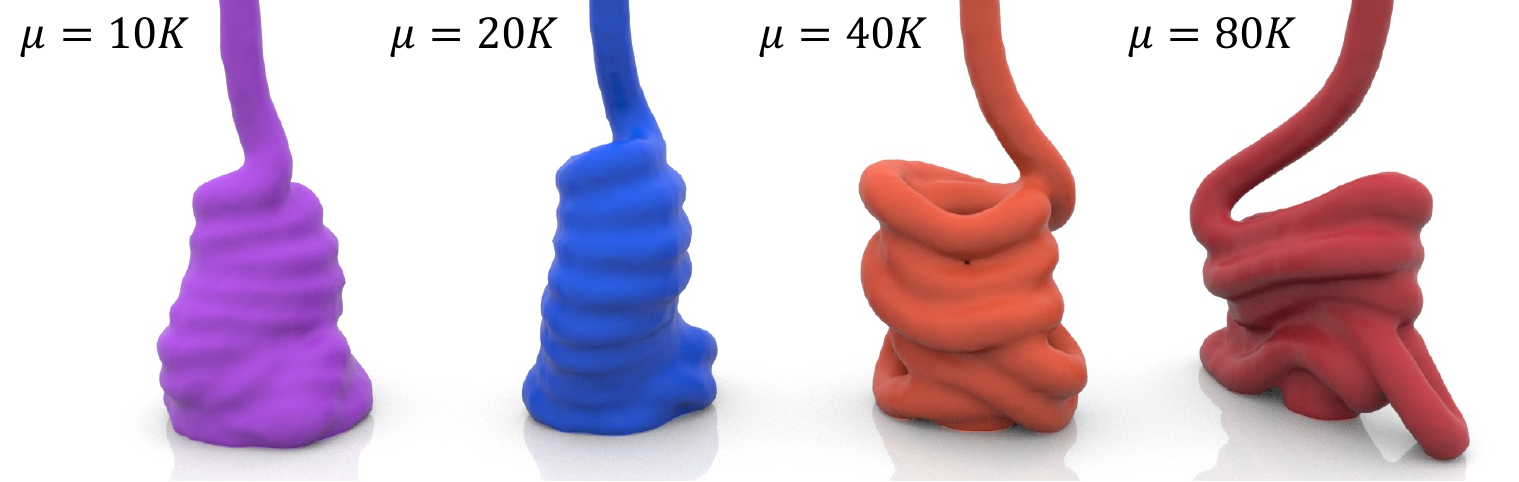}
  \caption{
XPBI effectively captures viscoplastic coiling; as the shear modulus $\mu$ increases from left to right, coils become more elastic.}
  \label{fig:viscoplastic}
\end{figure}

\begin{figure}[b]
\includegraphics[width=1.0\linewidth]{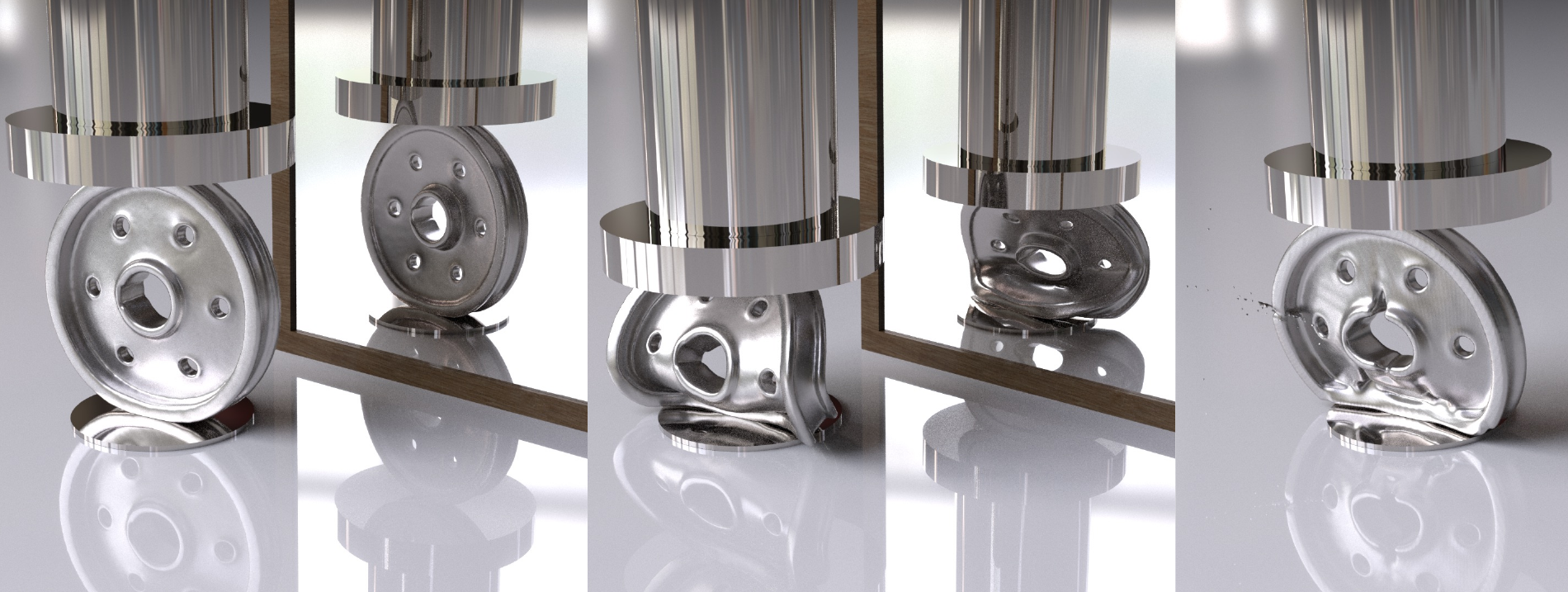}
  \caption{Hydraulic test on a stiff aluminum wheel: initial (\textbf{left}), w/ (\textbf{middle}) and w/o (\textbf{right}) position correction. The simulation suffers from artificial fracture and instability without our correction.}
  \label{fig:position_correction}
\end{figure}

\begin{algorithm}
\caption{Simulating $t^n \rightarrow t^{n+1}$ with XPBI}
\label{alg:cap}
\begin{algorithmic}
    \renewcommand{\COMMENT}[2][\linewidth]{\leavevmode\hfill\makebox[0pt][r]{\textcolor{black}{$\triangleright$
#2}}}
    \STATE Neighbor search using $\xx^n$ \COMMENT{\textsection~\ref{sec:neighbor}}
    \STATE Evaluate kernel gradient correction $\bm{L}_p$ \COMMENT{\textsection~\ref{sec:gradient}}
    \STATE $\bm{v} \leftarrow \bm{v}^n+\Delta t\MM^{-1}\ff_{\text{ext}}$
    \STATE $\bm{\lambda}=\bm{0}$
    
    \FOR{number of XPBD iterations}
        \FOR{all $p \in \{C_p\}$ looping with colored Gauss-Seidel\COMMENT{\textsection~\ref{sec:colored_gs}}}
            \STATE $(\nabla \bm{v})_p \leftarrow$ evaluateVelocityGradient$(\xx_p^n, \bm{v}_p)$ \COMMENT{\textsection~\ref{sec:gradient}}
            \STATE $\FF_p \leftarrow (\II+\Delta t (\nabla \bm{v})_p) \FF_p^n$
            \STATE $\FF_p \leftarrow \mathcal{Z}(\FF_p)$ \COMMENT{\textsection~\ref{sec:implicit_plasticity}}
            \IF{$C_p(\FF_p) \neq 0$}
                \STATE $\Delta\lambda_p = \frac{-C_p-\tilde{\alpha}\lambda}{\sum_{i=1}^N \frac{1}{m_i} |\nabla_{\xx_i}C_p(\xx)|^2 + \tilde{\alpha}}$\COMMENT{\textsection~\ref{sec:stvk}}
                \STATE $\lambda_p \leftarrow \lambda_p + \Delta \lambda_p$
                \STATE $\Delta\bm{v} = \frac{1}{\Delta t}\MM^{-1}\nabla C_p(\xx)^T \Delta \lambda_p$
                \STATE $\bm{v} \leftarrow \bm{v} + \Delta \bm{v}$
            \ENDIF
        \ENDFOR
        
        \FOR{all $i \in \{C_i\}$ looping with colored Gauss-Seidel}
            \STATE $\Delta\bm{v} = \frac{1}{\Delta t}\MM^{-1}\nabla C_i(\xx^n + \Delta t \bm{v})^T \Delta \lambda_i$ \small{(e.g., collision)}\COMMENT{\textsection~\ref{sec:xsph}}
            \STATE $\bm{v} \leftarrow \bm{v} + \Delta \bm{v}$
        \ENDFOR
    \ENDFOR
    
    \STATE $\bm{v}^{n+1} \leftarrow \bm{v}$
    \STATE Perform XSPH smoothing of $\bm{v}^{n+1}(\xx^n)$ \COMMENT{\textsection~\ref{sec:xsph}}
    \STATE Update $\FF^{n+1}$ and apply constitutive models \COMMENT{\textsection~\ref{sec:Fupdate}}
    \STATE $\xx^{n+1} \leftarrow \xx^n+ \Delta t \bm{v}^{n+1}$
    \STATE \footnotesize{\textbf{Note:} $\alpha=1/V_p^0$ and  $\tilde{\alpha}=\alpha/\Delta t^2$ for each constraint.}
\end{algorithmic}
\end{algorithm}

\subsection{Particle Neighbor Search} \label{sec:neighbor}

We reconstruct the neighbor information for each particle at the beginning of each timestep, similarly to \citet{macklin2013pbf}.  A  comprehensive overview of CPU- and GPU-based neighborhood search methods is surveyed by \citet{ihmsen2014sph}. Each material particle is assigned the same kernel radius in our discretization scheme. We adopt a uniform spatial-grid-based method for neighborhood searches following \citet{hoetzlein2014fast}. Particles are spatially stored in cells while neighbor lists $\mathcal{N}_p = \{ b \mid \|x_p - x_b\|_2 \leq k \}
$ are determined by querying adjacent grid cells using the Wendland kernel's support radius $k=2r$, and $r$ is the \ac{sph} particle kernel radius. In addition to the total number of particles $N=|\{p\}|$, the total number of particle neighbors $\sum |\mathcal{N}_p|$ is also critical for performance, as it determines the complexity of calculating inelastic constraints. We summarize the statistics for $\sum |\mathcal{N}_p|$ in Table.~\ref{table:timing}.

\begin{figure*} [t]
\includegraphics[width=1.0\linewidth]{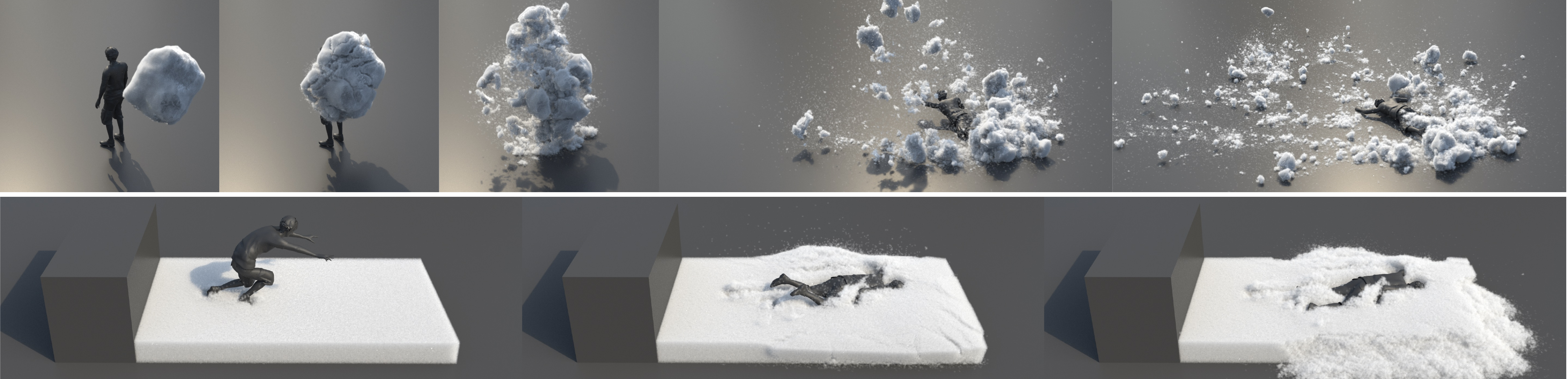}
  \caption{\textbf{Hitman and Snow Dive.} We successfully reproduce realistic and complex snow behaviors, such as a snowball hitting a person (\textbf{top}) and a person falling into a snow ground (\textbf{bottom}), using NACC \cite{wolper2019fracture} constitutive model.}
  \label{fig:snow}
\end{figure*}

\subsection{Colored Gauss-Seidel} \label{sec:colored_gs}

Original \ac{pbd} and \ac{xpbd} frameworks solve constraints iteratively using nonlinear Gauss-Seidel \cite{muller2007pbd, macklin2014unified, macklin2016xpbd}. In contrast, \citet{macklin2013pbf} adopted a Jacobi-style iteration for fluids, solving each constraint independently to enhance parallelism. We found that for high resolution and often high stiffness simulations considered in this paper, Jacobi iterations too slowly propagate information and often suffer from non-convergence (also noted by \citet{macklin2014unified}). Thus we implement colored Gauss-Seidel to maximize convergence, parallelism, and GPU throughput. We assign particles into cells with $\Delta x=2 r$ (\textsection~\ref{sec:neighbor}). $2^d$ colors are specified for eliminating dependencies between constraints in a $d$-dimensional simulation. We process all cells of the same color in parallel while constraints $C_p \in c_i$ corresponding to all particles in the same cell $c_i$ are computed serially.

The efficiency of this implementation heavily depends on the average \ac{ppc}, as particles within the same cell are traversed sequentially. To optimize \ac{ppc} and ensure an even particle distribution, we utilize Poisson disk sampling \cite{bridson2007fast} during the initial particle placement.

\subsection{XSPH and Position Correction} \label{sec:xsph}

The goal of \ac{xsph} is to incorporate artificial viscosity for mitigating nonphysical oscillations in dynamics observed in \ac{sph}-based simulations, which is more observable when the material is stiff and the constraints become hard to solve. We follow  \citet{schechter2012ghost}'s simpler \ac{xsph}-style  noise damping after velocity update by blending in surrounding particle velocities in $\Omega^n$:
\begin{align}
\bm{v}^{n+1}_p \leftarrow \bm{v}_p^{n+1} + c \sum_{b} V_b^n (\bm{v}_b^{n+1}-\bm{v}_p^{n+1}) W_b(\xx_p^n).
\end{align}
\ac{xsph} encourages smooth and coherent motion for inelastic materials in this paper. We use dimensionless $c=0.01$ in all examples.

In particle-based simulations, including but not limited to FLIP \cite{brackbill1986flip,zhu2005animating} and APIC \cite{jiang2015affine}, non-physical particle clumping and uneven particle distributions are common issues due to accumulation of advection errors. While \ac{mpm} is insensitive to the problem due to its structural grid nature, it is crucial for pure particle-based methods like \ac{sph} and our approach to maintain reasonably even particle distributions. Without doing so can strongly impair simulation quality and convergence, particularly when material stiffness is high. 

Various techniques in graphics have been proposed to address this by shifting particle positions \cite{ando2012preserving, ando2011particle, kugelstadt2019implicit}. However, we point out that while these methods work well for fluids, they are problematic for updated Lagrangian simulations because such positional shifting is transparent to the evolution of deformation gradients, leading to discrepancy between the positions of points and their perceived deformations. Fortunately within \ac{xpbd} we can directly adopt a point-point distance constraint \cite{macklin2014unified}
\begin{align}
C(\xx_p,\xx_b) = \left\| \xx_p - \xx_b\right\|_2 - r + \epsilon \geq 0
\end{align}
inside nonlinear iterations for all the neighborhood particle pairs. Here  
$r$ represents the particle kernel radius and $\epsilon$ is a small gap threshold used to determine when corrective action is needed for nearby particle pairs. We set $\epsilon$ to $0.25 r$. 

\subsection{Deformation Gradient Update} \label{sec:Fupdate}

Strain variables within XPBD iterations are temporary. After arriving at the post-XSPH velocity $\bm{v}^{n+1}$, we update both the deformation gradient state and position state using the same velocity -- an essential subtlety to maintain their consistency:
\begin{align}
{\FF}_p^{n+1} = \mathcal{Z}\left(\left(\II + \Delta t \frac{\partial \bm{v}^{n+1}}{\partial \xx}(\xx_p^n) \right) \FF_p^n\right),
\end{align}
where inelastic return mapping is also applied to ensure the stored elastic deformation gradient is within the yield region.

\begin{table*}[t]
\setlength{\tabcolsep}{4pt}
\centering
\caption{\textbf{Parameters and Statistics.} We summarize the parameters and timing statistics, including maximum particle numbers, the Wendland kernel radius, the average total number of particle neighbors per substep, the average time per frame, the \ac{xpbd} iterations per substep, and the time step size $\Delta t$ for various demos described in \textsection~\ref{sec:demos}. Material-related parameters are detailed in the last two columns. The abbreviations are NACC for Non-Associated Cam-Clay \cite{wolper2019fracture}, DP for Drucker-Prager \cite{klar2016sand}, VM for Von Mises \cite{li2022energetically}, and HB for Herschel-Bulkley \cite{yue2015continuum}. In addition to the basic settings of the material (density $\rho$, Youngs Modulus $E$, and Poisson Ratio $\nu$), we include model-specific parameters arranged as follows: 1) NACC: $(\rho, E, \nu, \alpha_0, \beta, \xi, M)$; 2) DP: $(\rho, E, \nu, \phi_f, c_0)$; 3) VM: $(\rho, E, \nu, \sigma_\gamma)$; and 4) HB: $(\rho, E, \nu, \sigma_\gamma, h, \eta)$. See references for detailed explanations of these parameters. 
}
\small{
\begin{tabular}{l|llrrlrrrr}
\hline
demo & particle \# & radius & ave  $\sum |\mathcal{N}_p|$ & ave sec/frame & iter \# & $\Delta t_{\text{frame}}$ & $\Delta t_{\text{step}}$ & material & material parameters \\
\hline
(Fig.~\ref{fig:noodles}) Noodles & 1.18M & $1/256$ & 28.7M & 46.3 & 10 & 1/40 & \(1 \times 10^{-4}\) & VM & \((1, 2 \times 10^{4}, 0.3, 76.9)\) \\
(Fig.~\ref{fig:cloth_sequence}) Cloth & 1.10M & $1/512$ & 19.9M & 24.3 & 10 & 1/100 & \(5 \times 10^{-5}\) & HB & \((100, 14754, 0.475, 50, 1, 10)\) \\
(Fig.~\ref{fig:camponotus}) Camponotus & 1.12M & $1/1024$ & 32.9M & 37.3 & 10 & 1/100 & \(4 \times 10^{-5}\) & NACC & \((2, 2\times 10^4, 0.35, -0.02, 0.5, 1, 2.36)\) \\
(Fig.~\ref{fig:castle}) Dam Breach & 4.00M & $1/384$ & 156.2M & 138.8 & 7 & 1/24 & \(2.5 \times 10^{-4}\) & DP & \((1, 400, 0.4, 30, 0.0007)\) \\
(Fig.~\ref{fig:hourglass}) Hourglass & 1.01M & $1/1024$ & 17.0M & 30.9 & 5 & 1/24 & \(1 \times 10^{-4}\) & DP & \((1, 3.537 \times 10^{5}, 0.3, 35, 0)\) \\
(Fig.~\ref{fig:snow}) Hitman & 1.05M & $1/512$ & 32.5M & 38.9 & 10 & 1/100 & \(4 \times 10^{-5}\) & NACC & $(4, 2 \times 10 ^ 4, 0.3, -0.005, 0.05, 30, 1.85)$ \\
(Fig.~\ref{fig:snow}) Snow Dive & 2.48M & $1/512$ & 64.1M & 78.2 & 5 & 1/100 & \(4 \times 10^{-5}\) & NACC & $(4, 1 \times 10 ^ 4, 0.3, -0.0005, 0.05, 30, 1.85)$ \\
(Fig.~\ref{fig:realtime}) Wrist & 20K & $1/256$ & 433.2K & 0.015 & 5 & 1/100 &  \(2 \times 10^{-4}\) & HB & \((100, 2250, 0.125, 10, 1, 10)\) \\
\hline
\end{tabular}}
\label{table:timing}
\end{table*}

\section{Results} 
\label{sec:results}

Here we evaluate and benchmark our \acf{xpbi} framework in terms of visual results against traditional \ac{xpbd} and \ac{mpm} methods, as detailed in \textsection~\ref{sec:evaluation}. Additionally, 
we present various demonstrations in \textsection~\ref{sec:demos} that illustrate XPBI's effective handling of diverse phenomena. We use Intel Core i9-14900KF CPU with 32GB memory and NVIDIA GeForce RTX 4090. We model common inelastic materials including Cam-Clay (\ac{nacc}) \cite{wolper2019fracture} snow and fracture, Drucker-Prager \cite{klar2016sand, tampubolon2017mixure} sand, Von Mises \cite{li2022energetically} plasticine and metal, and Herschel-Bulkley \cite{yue2015continuum} foam.

\subsection{Evaluation}
\label{sec:evaluation}

\begin{wrapfigure}{r}{0.5\linewidth}
\vspace{-1.5em}
 \hspace{-3.0em}
  \centering
  \includegraphics[width=1.2\linewidth]{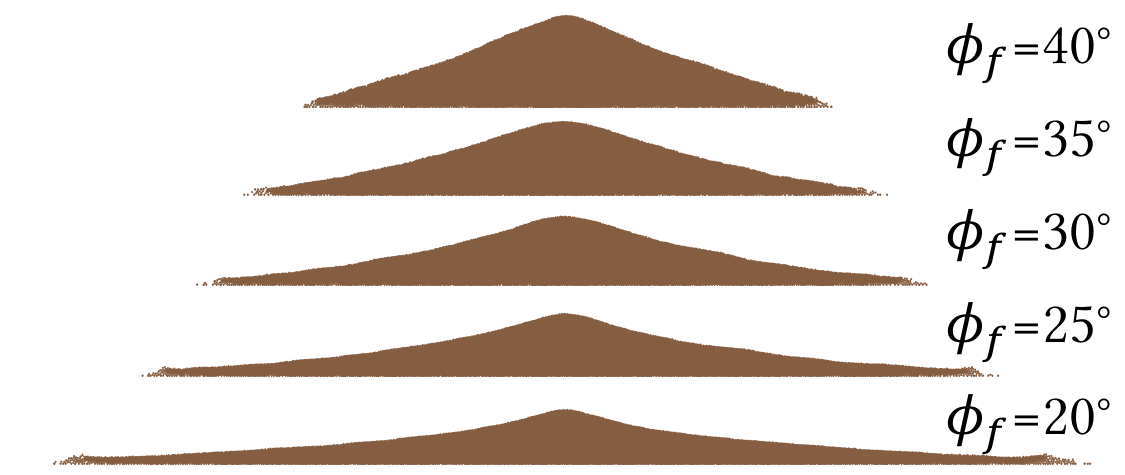}
  \vspace{-3.0em}
\end{wrapfigure}

\paragraph{Intuitive Parameters} We simulate sand collapsing with varying friction angles $\phi_f$. Our method reproduces characteristic piling shapes. In Fig.~\ref{fig:shear_thickening}, we compare viscoplastic, shear thinning, and shear thickening materials by only altering the Herschel-Bulkley power parameter $h$. Upon impact with a ground plane, the shear thickening material exhibits low flow rates under high stress, behaving elastically and bouncing off. Conversely, the shear thinning material flows immediately due to its higher flow rate. Similarly we can easily control the fluidity of viscoplastic goo (Fig.~\ref{fig:viscoplastic}). A smaller $\mu$ gives a more fluid-like appearance, while a larger $\mu$ leads to more elastic behavior.

\begin{figure} [b]
\includegraphics[width=1.0\linewidth]{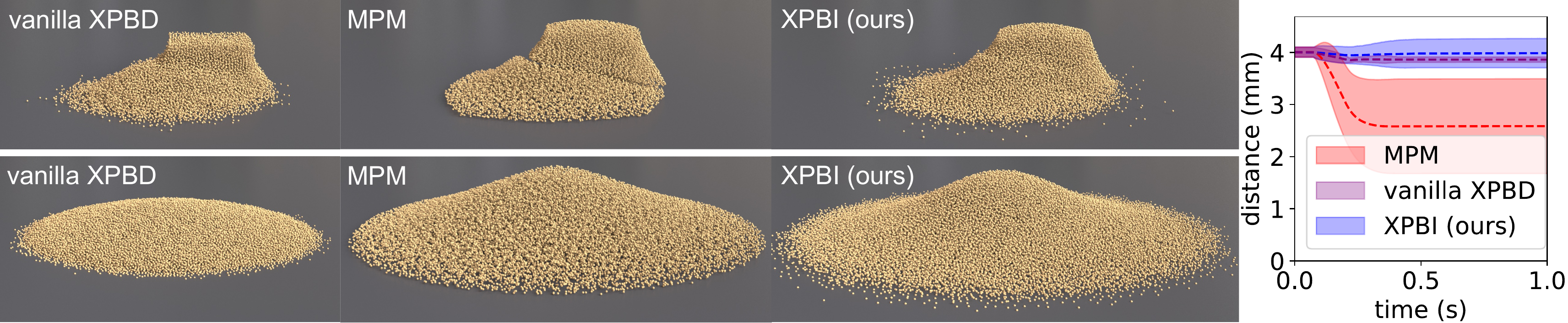}
  \caption{
  \textbf{Vanilla XPBD v.s. MPM v.s. XPBI.} We simulate \emph{two sand blocks collide} (\textbf{top}) and \emph{sand column collapse} (\textbf{bottom}) with vanilla XPBD \cite{macklin2014unified, macklin2016xpbd} (\textbf{left}), MPM \cite{klar2016sand} (\textbf{middle}), and XPBI (\textbf{right}).
} 
  \label{fig:mpm_pbd_pbi}
\end{figure}

\paragraph{Comparisons to Vanilla XPBD and MPM} We compare XPBI sand with both vanilla \ac{xpbd} \cite{macklin2014unified, macklin2016xpbd}, which employs a point-wise friction model, and explicit \ac{mpm} with Drucker-Prager plasticity \cite{klar2016sand}; see Fig.~\ref{fig:mpm_pbd_pbi}. This comparison includes simulations of \emph{two sand blocks collide} (top) and \emph{sand column collapse} (bottom). All methods apply a timestep of $\Delta t=0.1$ ms, with identical initial sampling positions for all particles across the methods. The vanilla \ac{xpbd} (left) approach fails to accurately replicate the correct friction angle upon sand settling. While both \ac{mpm} (middle) and XPBI (right) successfully model the continuum behavior of sand, our method achieves a more uniform particle distribution, avoiding the sparsity, clumping, and artificial grid $\Delta x$-gap phenomena typically caused by \ac{mpm} solvers. For a quantitative analysis, we also plotted the average distance between each particle and its nearest neighbor per frame, displayed on the far right of Fig.~\ref{fig:mpm_pbd_pbi}. Each particle was initially positioned at intervals equal to the particle radius. The average distance relative to the initial state in the \ac{mpm} decreases rapidly post-collision, accompanied by an increase in the variance of the distance, whereas our method maintains both the relative distance and variance stably throughout the simulation, indicating a more consistent particle distribution. It is also noteworthy that our method aligns more closely with the approach of \citet{yue2018hybrid} compared to \ac{mpm}, opting for \ac{dem} to capture more discrete behaviors near the free surface.

\paragraph{Implicit Plasticity} We evaluate our fully implicit plasticity treatment by replacing it with a semi-implicit method, which only performs return mappings at the end of the time steps as in \citet{stomakhin2013snow} \textsection~\ref{sec:Fupdate}), while \ac{xpbd} iterations only address elasticity. As shown in Fig.~\ref{fig:semi_implicit}, the semi-implicit approach (right) can lead to severe artifacts. This occurs because the forces generated by stresses outside the yield surface cause the continuum to behave more like a purely elastic body. This artifact results from the semi-implicit method's failure to account for plasticity during the \ac{xpbd} solve, leading to an overestimation of the material’s resistance to tensile deformation. In contrast, our method (middle) fully incorporates plasticity in the \ac{xpbd} iterations and avoids such artifacts.

\begin{figure} [tb]
\includegraphics[width=1.0\linewidth]{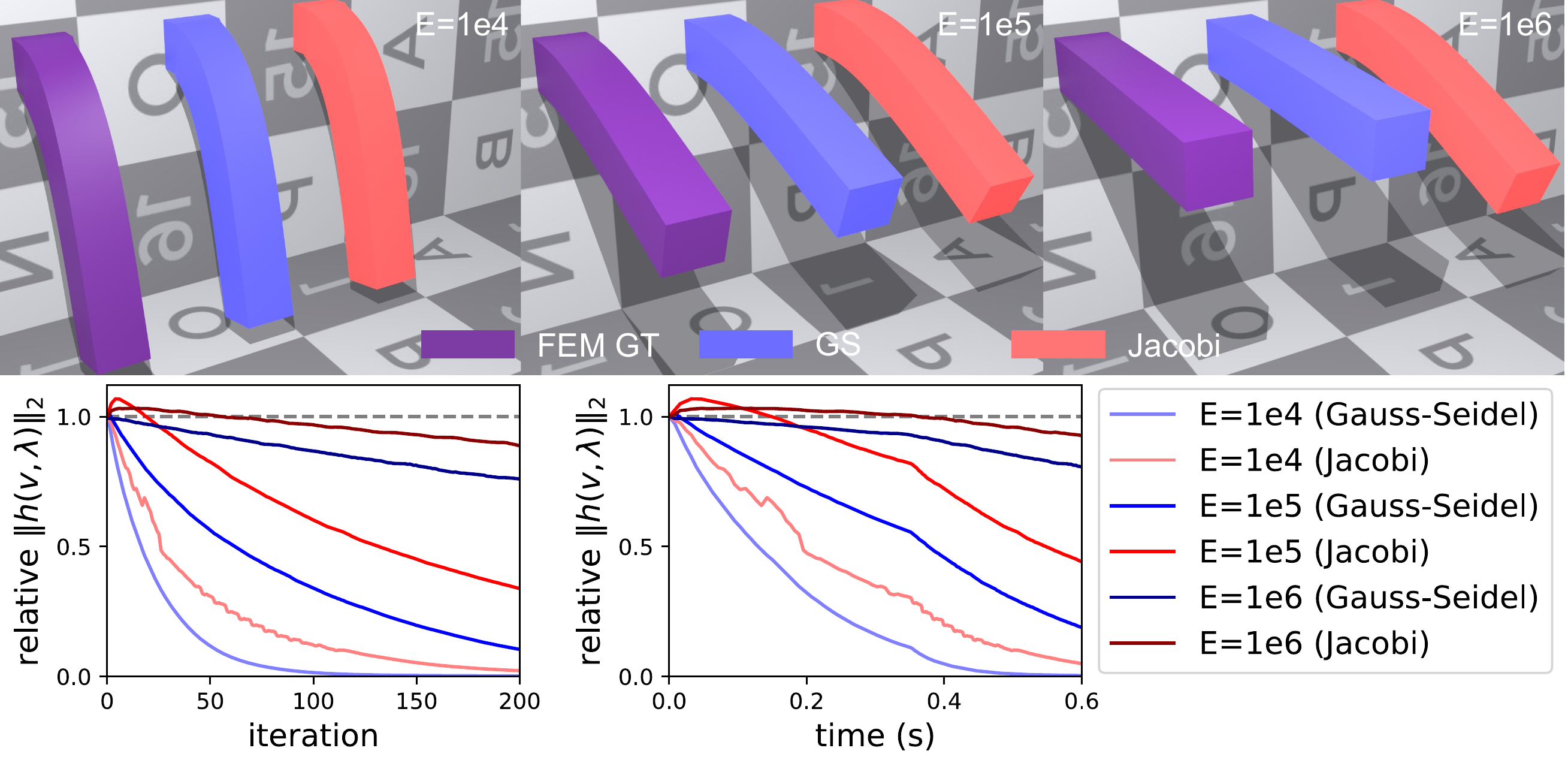}
  \caption{Cantilever beams modeled with StVK constitutive model using both our method and FEM ground truth with varying stiffness (\textbf{top}). Relative residual errors \(\bm{h(\bm{v},\bm{\lambda})}=\CC(\bm{v})+\tilde{\bm{a}} \bm{\lambda}\) with respect to iteration (\textbf{bottom left}) and runtime (\textbf{bottom right}) for a single frame from the above examples.}
  \label{fig:convergence}
\end{figure}

\paragraph{Convergence} We study the convergence of our method using cantilever beams of varying stiffness, $E=10^4\,\text{Pa}, 10^5\,\text{Pa}, 10^6\,\text{Pa}$, respectively (Fig.~\ref{fig:convergence}). We set the density at $100\,\text{kg/m}^3$ and the timestep at $\Delta t=5$\,ms. We also compare the relative residual errors of the Gauss-Seidel and Jacobi solvers in our method, as well as an implicit FEM ground truth, with respect to both iteration and time. XPBI with Gauss-Seidel can converge stably with a large timestep. In contrast, the Jacobi solver only converges with softer materials and struggles with high stiffness. This is consistent with observations about XPBD in prior work. Given that the materials discussed in our paper are predominantly very stiff, we opted for grid-colored Gauss-Seidel as our solver. Although using a large timestep is feasible with sufficient iterations, it becomes cost-inefficient if too many iterations are required. Thus, following \citet{macklin2019small}, we employ a small timestep.

\paragraph{Position Correction} To validate the importance of position correction, we conducted a hydraulic test on a highly stiff aluminum wheel, as shown in Fig.~\ref{fig:position_correction}. Without position correction, areas with significant deformation and stress suffer from gradient estimation errors due to uneven particle distribution, resulting in artificial fractures and eventually simulation instability. With position correction, however, we can reliably simulate high-stiffness materials under extensive deformation. This example also shows our capability in animating metal ductility using Von Mises plasticity.

\begin{figure} [t]
    \centering
\includegraphics[width=1.0\linewidth]{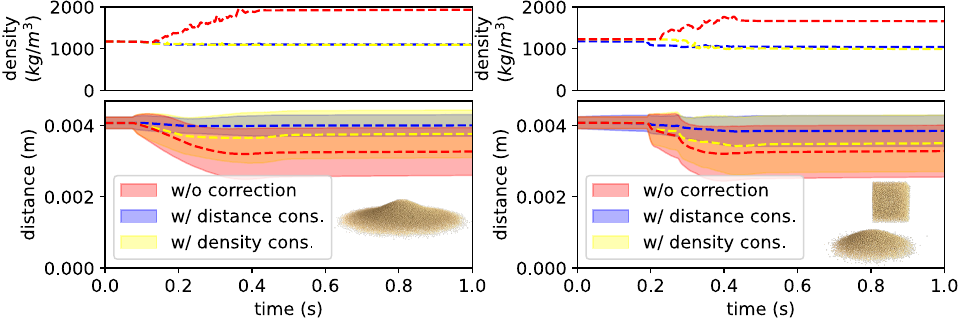}
  \caption{
Comparison of distance and density constraint. We plot the maximum density (\textbf{top}), the average distance to the nearest neighbor and respective standard deviations (\textbf{bottom}) for the settings described in Fig.~\ref{fig:mpm_pbd_pbi}.}
\label{fig:density_constraint}
\end{figure}

In addition to the distance constraint, other position correction strategies can also be employed. For instance, \citet{takahashi2019geometrically} demonstrated that the density constraint is effective in addressing particle clustering while preserving volume. In Fig.~\ref{fig:density_constraint}, we quantitatively compare the distance and density constraints using the same setup as in Fig.~\ref{fig:mpm_pbd_pbi}. Although both constraints are effective in maintaining maximum density during simulation, the distance constraint better resolves the distance between neighboring particles, improving both mean and variance, crucial for integration stability. We prefer the distance constraint for its simplicity.

\begin{figure} [b]
\includegraphics[width=1.0\linewidth]{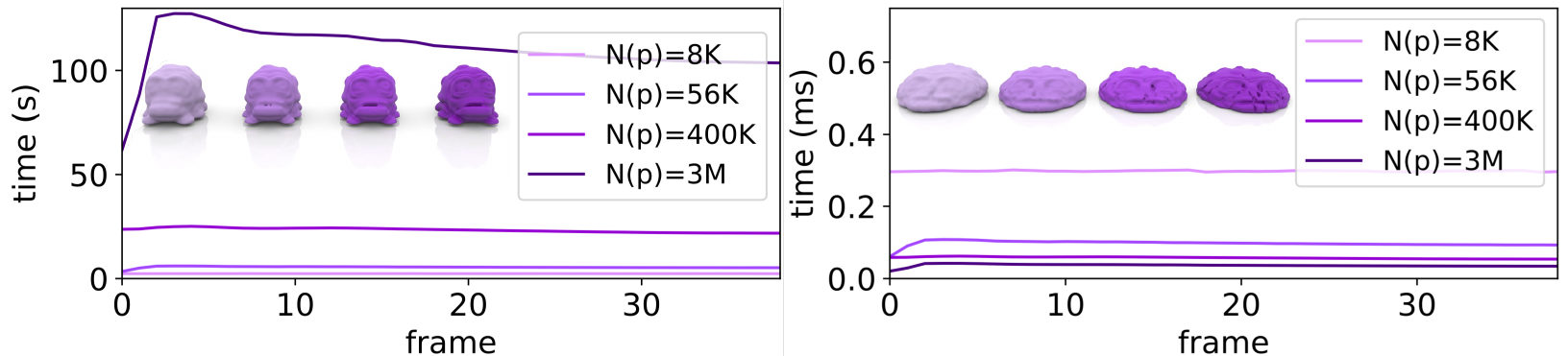}
  \caption{We simulate viscoplastic monsters falling to the ground using varying numbers of particles. We plot the computation time for each frame for 8K, 56K, 400K, and 3M (\textbf{left}) particles, demonstrating consistent behaviors across different particle counts. We also plot the average cost per particle (\textbf{right}). The running overhead of our algorithm decreases significantly as the number of particles increases, showing strong superlinear scalability.
  }
  \label{fig:particle_number}
\end{figure}

\paragraph{Scalability} To demonstrate the scalability of our method, we simulate viscoplastic monsters hitting the ground using 8K, 56K, 400K, and 3M particles, respectively. We maintain a constant simulation time step of $\Delta t = 0.1$ ms, perform 10 iterations of \ac{xpbd} per substep, and apply consistent material parameters across all simulations. Our method consistently replicates material behavior at varying resolutions, as depicted in Fig.~\ref{fig:particle_number}. We measure and plot the computation time for each individual frame (left) alongside the average computation time per particle (right). The average computation times per particle for 8K, 56K, 400K, and 3M are 0.30 ms, 0.098 ms, 0.058 ms, and 0.037 ms, respectively. These results highlight strong scalability; as the number of particles increases, the colored Gauss-Seidel solver can more effectively exploit GPU resources, significantly reducing the overall computational overhead per particle.

\begin{figure}[t]
    \centering
\includegraphics[width=1.0\linewidth]{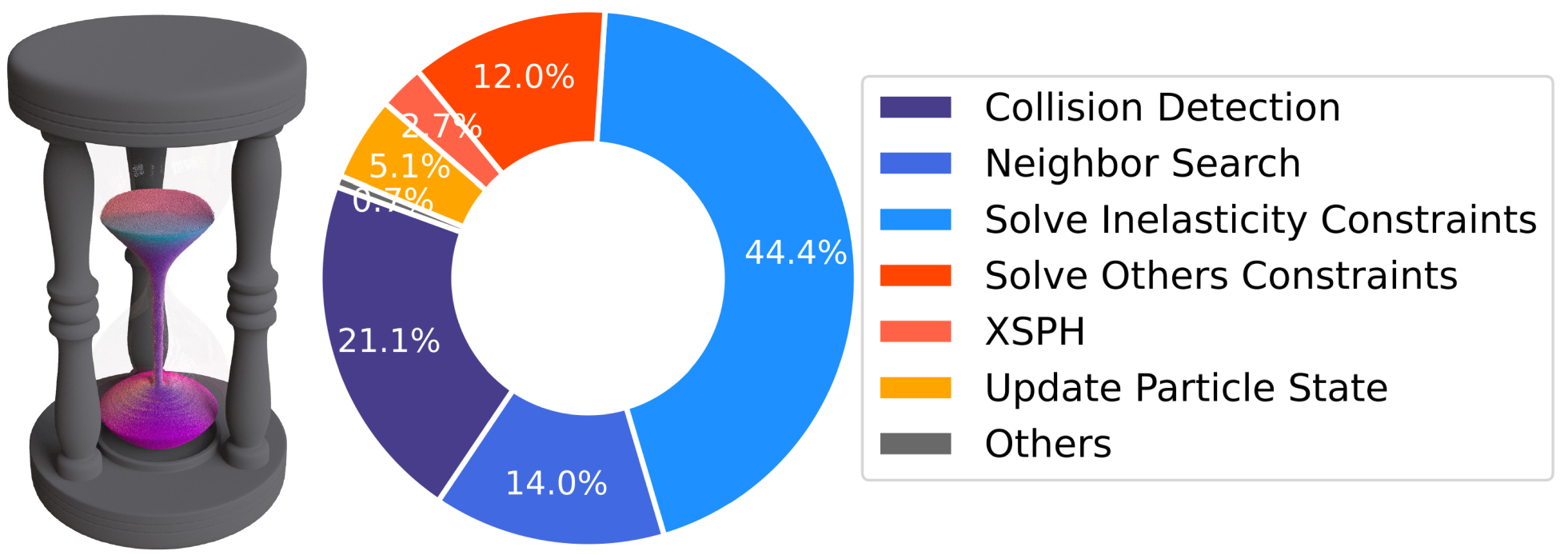}
  \caption{A typical breakdown of the total computational cost of our framework. We take the \emph{Hourglass} example (Fig.~\ref{fig:hourglass}) for demonstration.}
  \label{fig:breakdown}
\end{figure}

\paragraph{Timing Breakdown} Fig.~\ref{fig:breakdown} illustrates the GPU computational cost breakdown for the \emph{Hourglass} simulation example. In the breakdown, \emph{Collision Detection} refers to the time spent on constructing the LBVH \cite{karras2012maximizing} and querying for collision between point-triangle pairs. \emph{Neighbor Search} covers the time taken to build the background grid and neighborhood list by querying adjacent cells (see \textsection~\ref{sec:neighbor} for details). \emph{Solve Inelasticity Constraints} involves our colored Gauss-Seidel solver for inelasticity constraints, including fixed-point implicit plasticity treatment. \emph{Solve Other Constraints} accounts for the time spent on resolving all other constraints, such as point-triangle distance constraints for boundary conditions, position correction, as well as stretching, bending, and density constraints in other examples. \emph{\ac{xsph}} and \emph{Update Particle State} are detailed in \textsection~\ref{sec:xsph} and \textsection~\ref{sec:Fupdate}, respectively. The majority of our framework’s computational time is spent on solving inelasticity constraints, while additional stability enhancements like \ac{xsph} and position correction contribute a relatively small overhead.

\begin{figure}[b]
 \centering
 \includegraphics[width=1.0\linewidth]{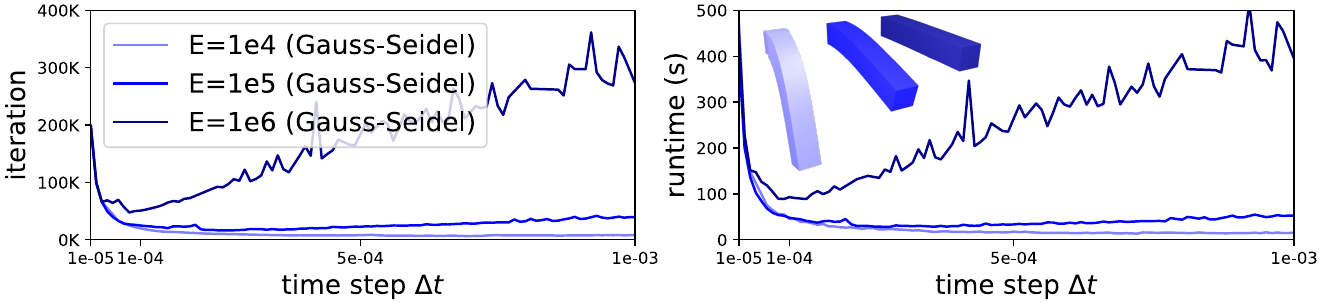}
  \caption{Ablation on varying time step sizes. Total XPBD iterations 
 (\textbf{left}) and runtime (\textbf{right}) required for all steps to converge to a fixed residual error for 1s simulation time with varying time step sizes.}
  \label{fig:dt_ablation}
\end{figure}

\paragraph{Choice of Time Step} While our method with colored GS solver can converge with a time step $5\times$ larger, as noted by \citet{macklin2019small}, smaller time steps are generally preferable due to the nonlinear increase in GS iterations needed for convergence. However, we emphasize that reducing to just one iteration per time step is sub-optimal in our scenario, as shown in Fig.~\ref{fig:breakdown}, where collision detection, neighbor search, XSPH, and particle state updates occur once per time step, accounting for about 1/3 of the total computation time. A timestep that is too small increases the overhead of these operations, with minimal benefit to GS convergence. We study the efficiency of different time step sizes by applying the same setting as in Fig.~\ref{fig:convergence}, this time fixing the residual error threshold $\epsilon_E$ and ensuring that each timestep converges under the given threshold, $\| h(v, \lambda) \|_2 \leq \epsilon_E$, with adaptive XPBD iterations. The $\epsilon_E$ values are chosen based on the residual errors observed when the cantilever beams exhibit visually identical behaviors to the FEM ground truth for each stiffness $E$, respectively. We measure the total XPBD iterations and runtime required relative to the timestep size for a 1-second simulation. As shown in Fig.~\ref{fig:dt_ablation}, when stiffness is high, the total number of iterations required for convergence increases with the timestep size. Interestingly, when the timestep is sufficiently small, the total XPBD iterations required actually increases, as each timestep necessitates at least one iteration, and the runtime's growth rate rises further due to the additional overhead per timestep. In practice, we select a $\Delta t$ between $5\times{10}^{-5}\,\text{s}$ and $2\times{10}^{-4}\,\text{s}$.

\begin{figure} [b]
    \centering
\includegraphics[width=1.0\linewidth]{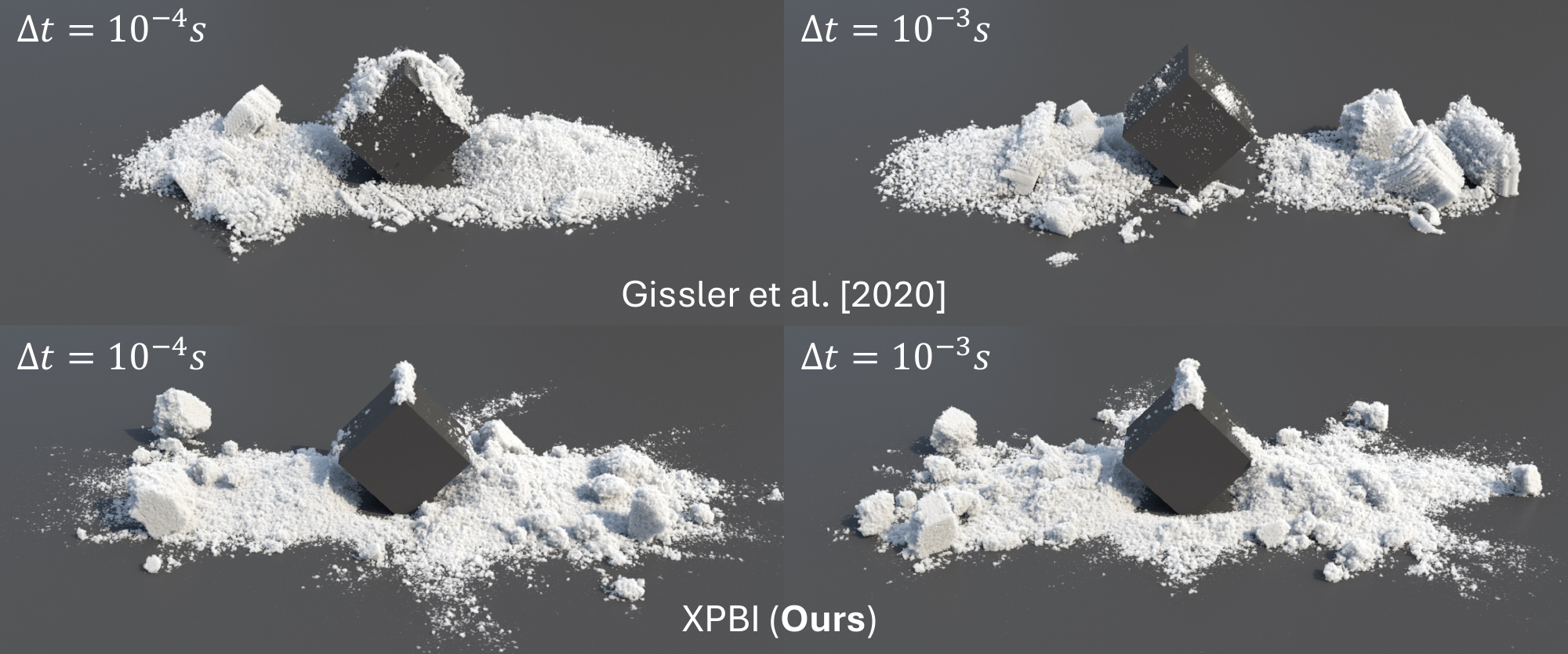}
  \caption{Comparison with \citet{gissler2020implicit}. The semi-implicit plasticity approach in \citet{gissler2020implicit} (\textbf{top}) exhibits timestep-dependent behavior, whereas our method (\textbf{bottom}) demonstrates consistent behavior across different timestep sizes.}
  \label{fig:sph_snow}
\end{figure}

\paragraph{Comparison with \citet{gissler2020implicit}} Our method shares similarities with \citet{gissler2020implicit} in computing the velocity gradient and advecting the deformation gradient using SPH-based spatial discretization. However, \citet{gissler2020implicit} utilizes a Jacobi solver for the primal system, whereas we solve the dual formulation with a Lagrangian multiplier, enabling coupling with traditional PBD materials. A comprehensive discussion of the advantages/disadvantages of dual vs. primal formulations can be found at \citet{macklin2020primal}. Most notably, as shown in Fig.~\ref{fig:sph_snow}, our approach to plasticity is distinct. We adopt the same snow constitutive model \shortcite{stomakhin2013snow} for both methods, with parameters: $\rho=400\,\text{kg/m}^3$, $E=5\times10^5\,\text{Pa}$, hardening coefficient $\xi=10$, critical compression $\theta_c=0.025$, and critical stretch $\theta_s=0.0075$. We conducted simulations using timesteps of $\Delta t={10}^{-4}\,\text{s}$ and ${10}^{-3}\,\text{s}$. \citet{gissler2020implicit} employs a semi-implicit plasticity model with a single post-return-mapping projection per time step, akin to \citet{stomakhin2013snow}, making snow behavior timestep dependent due to this explicit plastic deformation update. In contrast, our method uses a fully implicit plasticity treatment, alternating between XPBD iteration and fixed-point iteration and producing consistent behaviors across different timestep sizes.

\subsection{Demos}
\label{sec:demos}

Complex materials with up to millions of particles, such as mud (Fig.~\ref{fig:noodles}, Fig.~\ref{fig:castle}), viscoplastic paint (and its coupling with traditional XPBD cloth) (Fig.~\ref{fig:cloth_sequence}), brittle fracture (Fig.~\ref{fig:camponotus}), sand (Fig.~\ref{fig:hourglass}), and snow (Fig.~\ref{fig:snow}) can be simulated with XPBI. The timing and parameters are summarized in Table.~\ref{table:timing}.

\begin{figure} [t]
\includegraphics[width=1.0\linewidth]{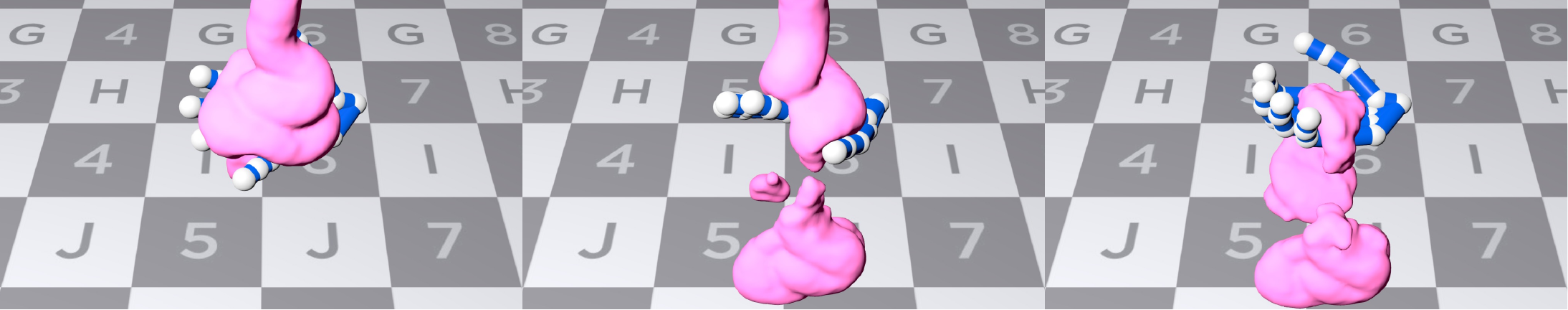}
  \caption{\textbf{Real-time Vision Pro\textsuperscript{TM} Interaction.} Interactive manipulation of a viscoelastic fluid, consisting of up to 20K particles, is simulated at 30fps.}
\label{fig:realtime}
\end{figure}

\paragraph{Real-time Interaction} The position-based method family is widely adopted in game and VR applications due to its real-time interactive capabilities \cite{barreiro2017conformation, jiang2024vr}. We further showcase our method in interactive applications where very small time steps are impractical. The convergence and stability of our approach enable interactive performance in moderately complex scenarios involving 20K particles. For this application, we employ the Jacobi solver due to its significant parallelism capabilities on GPUs for small-scale simulation. We use an Apple Vision Pro VR device and \emph{VisionProTeleop} \cite{park2024avp} to track hand motions and enable a virtual hand to interact with viscous fluids.

\section{Discussion}

In summary, XPBI is a novel updated Lagrangian enhancement for \ac{xpbd}, enhancing its capability for simulating complex inelastic behaviors governed by continuum mechanics-based constitutive laws. Further incorporating an implicit plasticity treatment and stability enhancements, XPBI can be easily integrated into standard \ac{xpbd} to open up its new simulation possibilities.

Given the high stiffness and detailed resolution of most scenes, the timestep is constrained by the relative low GS convergence and SPH CFL condition, which is related to the kernel's support radius. We note that although our method falls under the category of implicit methods capable of handling highly stiff materials, additional damping models, such as XSPH or XPBD constraint damping, are still necessary to avoid jittering effects or stability issues.

While XPBI supports a broad range of material behaviors, Maxwell viscoelastic materials \cite{fang2019silly} necessitate more specialized treatment. Also, interactions between sand and water mixtures occur primarily at the material boundary. Simulating actual porous media \cite{tampubolon2017mixure} and fluid sediment mixture \cite{gao2018animating} with proper momentum transfer are interesting future work. Another direction is to optimize parallelism for solving inelastic per-particle constraints. Although grid-colored Gauss-Seidel significantly improves performance over sequential iterations in large-scale simulations, it underperforms in small-scale cases on modern GPUs due to low utilization, obstructing many interactive rate experiments. A tailored  real-time solver would be interesting.

\begin{acks}
We thank the anonymous reviewers for their valuable feedback. We thank Siyu Ma for contributing to the demo design discussions. We acknowledge support from NSF (2301040, 2008915, 2244651, 2008564, 2153851, 2023780), UC-MRPI, Sony, Amazon, and TRI.
\end{acks}

\bibliographystyle{ACM-Reference-Format}
\bibliography{reference}

\end{document}